\documentclass[journal]{IEEEtran}
\usepackage[dvips]{graphicx}
\usepackage{mathrsfs}
\usepackage{amssymb}
\usepackage{latexsym}
\usepackage{multirow}
\usepackage[cmex10]{amsmath}
\usepackage[caption=false]{subfig}

\newtheorem{theorem}{Theorem}

\newtheorem{corollary}{Corollary}

\newtheorem{definition}{Definition}

\newtheorem{lemma}{Lemma}

\newtheorem{proposition}{Proposition}

\newcommand{\beqn}{\begin{equation}}
\newcommand{\eeqn}{\end{equation}}
\newcommand{\beqa}{\begin{eqnarray}}
\newcommand{\eeqa}{\end{eqnarray}}
\newcommand{\beqas}{\begin{eqnarray*}}
\newcommand{\eeqas}{\end{eqnarray*}}

\begin{document}
\title{Near-Optimal Deviation-Proof Medium Access Control Designs in Wireless Networks}
\author{{Khoa Tran Phan, Jaeok Park, and Mihaela van der
Schaar}
\thanks{The authors are with Electrical Engineering Department, University of
California, Los Angeles (UCLA), 420 Westwood Plaza,
Los Angeles, CA 90095-1594, USA. e-mail: \{kphan, jaeok, mihaela\}@ee.ucla.edu.}
}
\maketitle
\markboth{PHAN, PARK, AND VAN DER SCHAAR: Near-Optimal Deviation-Proof Medium Access Control Designs IN WIRELESS NETWORKS}{}
\begin{abstract}
Distributed medium access control (MAC) protocols are essential for the
proliferation of low cost, decentralized wireless local area networks (WLANs).
Most MAC protocols are designed with the presumption that nodes comply with
prescribed rules.
However, selfish nodes have natural motives to manipulate protocols in order to improve their own performance. This 
often degrades the performance of other nodes as well as that of the overall system.
In this work, we propose a class of protocols
that limit the performance gain which nodes can obtain through selfish manipulation while incurring only
a small efficiency loss.
The proposed protocols are based on the idea of a
review strategy, with which nodes collect signals about the actions of other nodes
over a period of time, use a statistical test to infer whether or not
other nodes are following the prescribed protocol, and trigger a punishment
if a departure from the protocol is perceived. We consider the cases of private
and public signals and provide analytical and numerical results to demonstrate the
properties of the proposed protocols.
\end{abstract}


\section{Introduction}

\IEEEPARstart{I}{n wireless} communication networks, multiple nodes often
share a common channel and contend for access. To resolve
contention among nodes, many different MAC protocols have been devised and are currently used in international standards
(e.g., IEEE 802.11a/b/g protocols) \cite{WLAN}.
When a MAC protocol is designed, two types of node behavior can be assumed.
One is \emph{cooperative} nodes that comply with prescribed protocols, and the other
is \emph{selfish} nodes that are capable of manipulating prescribed protocols in order to improve their
own performance.
With cooperative nodes, a MAC protocol can be designed to optimize
the system performance \cite{Chen}--\cite{Azouzi}.
However, such a protocol is not robust to selfish manipulation
in that selfish nodes that can re-configure the software or firmware
may want to deviate from the protocol in pursuit of their self-interest \cite{HubauxBook}.
Thus, selfish manipulation often results in a suboptimal outcome, different
from the one desired by the protocol designer \cite{Tan}--\cite{Kyasanur}.
On the other hand, a MAC protocol can be designed assuming selfish nodes
so that the protocol is \emph{deviation-proof} in the sense that
selfish nodes do not find it profitable to deviate from the protocol.
However, the incentive constraints imposed by the presence of
selfish nodes in general restrict the system performance
\cite{Altman1}, \cite{Azouzi}. In this paper, we aim to
resolve the tension between the selfish manipulation and optimal performance
by proposing a class of slotted MAC protocols that limit the performance gain from selfish manipulation
while incurring only a small efficiency loss compared to the optimal performance achievable with cooperative nodes.

Recently, a variety of slotted MAC protocols have
been designed and analyzed using a game theoretic framework.
With cooperative nodes, protocols can be designed to achieve system-wide optimal outcomes.
In \cite{Park}, a class of slotted MAC protocols is proposed in which nodes can self-coordinate their transmission slots
based on their past transmission actions and feedback information to achieve a time division multiple access (TDMA) outcome.
In \cite{Ma}, the authors propose generalized slotted Aloha protocols that maximize system throughput given a
short-term fairness constraint. In \cite{Altman1}, \cite{Azouzi}, variations of slotted MAC
protocols with different capture effects, prioritization, and power diversity are studied.
It has been demonstrated that with cooperative nodes one can obtain optimal throughput and
expected delay as well as system stability.

Selfish behavior in MAC protocols has also been analyzed using game theory.
In \cite{MacKenzie}, the authors establish the stability region for a slotted
Aloha system with multipacket reception and selfish nodes. In \cite{Jin1}, the authors
study the existence of and convergence to Nash equilibrium in a slotted Aloha
system where selfish nodes have quality-of-service requirements.
It is often observed that selfish behavior often leads to suboptimal outcomes.
For example,
a prisoners' dilemma phenomenon arises among
selfish nodes using the generalized slotted Aloha protocols of \cite{Ma}.
A decrease in system throughput,
especially when the workload increases due to the selfish behavior of nodes, is observed in \cite{Altman1}, \cite{Azouzi}.
In the 802.11 distributed MAC protocol, competition among selfish nodes results in an
inefficient use of the shared channel in Nash equilibria \cite{Tan}.

Research efforts have been made to devise MAC protocols that sustain optimal outcomes among selfish nodes.
In \cite{Park1}, the authors induce selfish nodes to behave cooperatively in a slotted random access network by introducing an intervening
node that monitors the actions of nodes and decides its intervention level accordingly.
Pricing has also been used as a method to incentivize selfish nodes.
In \cite{Altman1}, the authors avoid the degradation of system throughput due to selfish behavior by adding a cost
of transmissions and retransmissions. In \cite{Jin}, the network charges
nodes for each successfully transmitted packet, and the authors consider
the problem of adjusting the price-per-packet to achieve a desired operating point.
The above approaches, however, require a central entity, which may not be available
in a distributed environment. In the case of an intervention mechanism,
an intervening node that is capable of monitoring and intervening should be present in the system.
In the case of a pricing mechanism, a billing authority is needed to charge
payments depending on the usage of the network.
In this paper, we propose an approach that decentralizes into nodes the burden of
monitoring and punishing.


To this end, we rely on the theory of repeated games \cite{Mailath} to sustain cooperation
among selfish nodes. When the nodes in a system interact repeatedly, they can
make their decisions dependent on their past observations. Thus, nodes
can trigger a punishment when they observe a deviation from a predetermined
operating point. If the loss due to punishment outweighs the gain from deviation,
selfish nodes do not have an incentive to deviate from a predetermined
operating point.
The idea of using a repeated game strategy to build a deviation-proof protocol
has recently been applied to several problems in communications and networking (see, for example, \cite{Wu}--\cite{Pandana}).
However, most existing work assumes perfect monitoring, where players observe
decisions that other players make. With perfect monitoring, it is
relatively easy to construct a deviation-proof protocol by using a trigger
strategy, which is commonly used to prove various versions of the Folk theorem.

In our work, we consider a scenario where the decisions of nodes are
their transmission probabilities, which cannot be observed directly.
In order to design a deviation-proof protocol, we use the idea of
a review strategy \cite{RoyRadner}, \cite{RoyRadner1} with which
nodes collect imperfect signals about the decisions of other nodes,
perform a statistical test to determine whether or not a deviation has occurred,
and trigger a punishment if they conclude so.
Our main contributions in this paper can be summarized as follows.
\begin{itemize}
\item We model a slotted multiple access communications scenario
as a repeated game, which allows us to adopt a repeated game strategy, including
a review strategy, to design a protocol.
\item We first consider the case where nodes observe private signals on the channel access outcomes.
We design deviation-proof protocols assuming that a deviating node can employ only a deviation
strategy using a constant transmission probability.
We provide a necessary and sufficient condition for a given protocol to be deviation-proof.
We show that the efficiency loss of a deviation-proof protocol can be made arbitrarily small
if there is a statistical test that becomes perfect as more signals are accumulated.
\item We also consider the case where nodes observe public signals on the channel access outcomes.
We show that with public signals it is possible to design near-optimal deviation-proof protocols 
even when nodes can use any deviation strategy.
\item Besides slotted MAC protocols, we provide a possible application of our design methodology to the case of CSMA/CA protocols with selfish nodes.
\item We illustrate the properties of the proposed protocols with numerical results.
\end{itemize}
The proposed protocols are fully distributed in the sense that they need no central entity
to coordinate the operation of nodes and that nodes take actions depending solely
on their own local information without communicating with other nodes.

The rest of this paper is organized as follows. In Section II, we formulate a repeated game model
for slotted multiple access communications. In Section III, we propose and analyze
deviation-proof protocols based on a review strategy when signals are private, with an example presented in Section IV.
In Section V, we investigate deviation-proof protocols when signals are public, with an example presented in Section VI.
In Section VII, we discuss a possible extension of the proposed protocols to a CSMA/CA network with selfish nodes.
We conclude the paper in Section VIII.

\section{Repeated Game Framework for Slotted Multiple Access Communications}

\subsection{Stage Game} \label{gamemodel}

We consider a wireless communication network with a set $\mathcal{N} = \{1,2,\hdots,N\}$ of
$N$ nodes interacting over time. Time is divided into slots of equal length, and in each slot, a node has a packet to
transmit (i.e., saturated arrivals) and can attempt to send the packet or wait. Due to interference in the shared
communication channel, a packet is transmitted successfully only if there is no other packet
transmitted in the same slot. If more than one transmission takes place in a slot, a collision occurs and no packet is
transmitted successfully. We model the interaction of nodes in a single slot
as a non-cooperative game in normal form, called the \emph{random access game}.

The set of pure actions available to node $i \in \mathcal{N}$ in a slot is $A_i
\triangleq \{T,W\}$, where $T$ stands for ``transmit'' and $W$ for
``wait.'' We denote the pure action of node $i$ by $a_i \in A_i$ and a pure action profile
by $\mathbf{a} \triangleq (a_1,\ldots,a_N) \in \mathcal{A} \triangleq \prod_{i \in \mathcal{N}} A_i$.
A mixed action for node~$i$ is a probability distribution on $A_i$. Since there are only two pure actions, a
mixed action for node~$i$ can be represented by a transmission probability $p_i \in [0,1]$, and the
set of mixed actions for node $i$ can be written as $P_i \triangleq [0,1]$. A mixed action profile is
denoted by $\mathbf{p} \triangleq (p_1,\ldots,p_N) \in \mathcal{P} \triangleq \prod_{i \in \mathcal{N}}P_i$.
The payoff function of node $i$ is defined by $u_i:\mathcal{A} \rightarrow \mathbb{R}$, where
$u_i(\mathbf{a}) = 1$ if $a_i = T$ and $a_j = W$ for all $j \neq i$ and $u_i(\mathbf{a}) = 0$
otherwise. That is, a node receives payoff 1 if it has a successful transmission and 0 otherwise.
Then, the expected payoff of a node is given by the probability that it has a successful transmission,
and with a slight abuse of notation, the payoff of node $i$ when mixed action profile $\mathbf{p}$ is chosen can be written as
\beqn 
u_i(\mathbf{p}) = p_i \prod_{j \in \mathcal{N} \setminus \{i\}} (1-p_j). \nonumber
\eeqn

The random access game is defined by the tuple $\Gamma \triangleq \left\langle \mathcal{N}, (A_i)_{i \in \mathcal{N}}, (u_i)_{i \in \mathcal{N}} \right\rangle$.
It is well-known from the static analysis of the random access game that there is at least
one node~$i$ choosing $p_i=1$ at any pure strategy Nash equilibrium (NE) \cite{Cagalj}, \cite{Park1}.
That is, when nodes myopically maximize their own payoffs, there is at least one node always
transmitting its packets, and thus there can be at most one node obtaining a positive payoff.
Moreover, in the unique symmetric NE, every node transmits with probability 1, which
results in zero payoff for every node.
On the other hand, the symmetric Pareto optimal (PO) outcome is achieved when each node chooses $p_c = 1/N$, which
yields a positive payoff $u^{\rm PO} = (1-1/N)^{N-1}/N$ for every node \cite{massey}. We call $p_c$ the \emph{cooperation
probability} and $u^{\rm PO}$ the optimal payoff.

\subsection{Repeated Game}

We now formulate the repeated random access game, where the actions of a node
can depend on its past observations, or information histories.
Time slots are indexed by $t = 1,2,\ldots$. At the end of each slot, nodes obtain signals on
the pure action profile chosen in the slot. Let $Z_i$ be the finite set of signals
that node $i$ can receive. Define $\mathcal{Z} \triangleq \prod_{i \in \mathcal{N}} Z_i$,
and let $Q$ be a mapping from $\mathcal{A}$ to $\Delta(\mathcal{Z})$, where $Q(\mathbf{a})$ represents
the distribution of signals when nodes choose pure action profile $\mathbf{a}$.
A \emph{signal structure} is specified by the pair $(\mathcal{Z}, Q)$.
We say that signals are \emph{private} if there exist $\mathbf{z} \triangleq (z_1, \ldots, z_N) \in \mathcal{Z}$
and $\mathbf{a} \in \mathcal{A}$ such that $z_i \neq z_j$ for some $i, j \in \mathcal{N}$
and $\mathbf{z}$ occurs with positive probability in $Q(\mathbf{a})$.
We say that signals are \emph{public} if they are not private.
That is, signals are private if it is possible
for nodes to receive different signals, whereas signals are public if signal realization is the same
for all nodes.

The history of node $i$ in slot $t$, denoted by $h_i^t$, contains the signals that node $i$ has received by the
end of slot $t-1$. That is, $h_i^t = (z_i^0,\hdots,z_i^{t-1})$, for $t=1,2,\ldots$, where
$z_i^t$ represents the signal that node $i$ receives in slot $t$ and $z_i^0$ is set as
an arbitrary element of $Z_i$.\footnote{In slot $t \geq 2$, node $i$ also knows its past mixed actions
$(p_i^1,\hdots,p_i^{t-1})$ and their realizations $(a_i^1,\hdots,a_i^{t-1})$.
However, since we focus on repeated game strategies using only past signals, we do not include them in our
history specification.}
The set of slot $t$ histories of node $i$ is written as $H_i^t$, and the set of all possible
histories of node $i$ is given by $H_i \triangleq  {\cup}_{t = 1}^{\infty} H_i^t$.
The (behavior) strategy of node $i$ specifies a mixed action for node $i$ in the stage game conditional
on a history it reaches. Thus, it can be represented by a mapping
$\sigma_i: H_i \rightarrow P_i$. We use $\Sigma_i$ to denote the set of strategies
of node $i$. We define a \emph{protocol} as a strategy
profile $\sigma \triangleq (\sigma_1, \ldots, \sigma_N) \in \Sigma \triangleq \prod_{i \in \mathcal{N}} \Sigma_i$.
To evaluate payoffs in the repeated game model, we use the limit of means criterion since the length
of a slot is typically short.\footnote{For example, the slot duration of the 802.11 DCF basic access method is 20$\mu$s \cite{WLAN}.},
A protocol $\sigma$ induces a probability distribution on the
sequences of mixed action profiles $\{\mathbf{p}^t\}_{t=1}^{\infty}$,
where $\mathbf{p}^t$ is the mixed action profile in slot $t$. The payoff of node $i$ under protocol $\sigma$
can be expressed as
\beqn
U_i(\sigma) = \lim_{J \rightarrow \infty} E\left[\frac{1}{J}\sum_{t=1}^J u_i(\mathbf{p}^t) \bigg| \sigma \right], \nonumber 
\eeqn
assuming that the limit exists. If the limit does not exist, we replace the operator $\lim$ by $\liminf$.\footnote{Although we consider
the limit of means criterion, the following results can be extended with a complication to the case of the discounting criterion
as long as the discount factor is close to 1, as in \cite{RoyRadner}.
Our analysis can also be extended to the case where a node incurs a transmission cost
whenever it attempts transmission, as long as the cost is small.}

We say that a signal structure $(\mathcal{Z}, Q)$ is symmetric if $Z_1 = \cdots = Z_N$
and the signal distribution $Q$ is preserved under permutations of indices for nodes.
With a symmetric signal structure, we have $H_1 = \cdots = H_N$ and thus
$\Sigma_1 = \cdots = \Sigma_N$ since $P_1 = \cdots = P_N$.
We say that a protocol $\sigma$ is symmetric if it prescribes
the same strategy to every node, i.e., $\sigma_1 = \cdots = \sigma_N$.
In the remainder of this paper, we assume that the signal structure is symmetric
and focus on symmetric protocols. Since a symmetric protocol can be represented with a strategy,
we use the two terms ``protocol'' and ``strategy'' interchangeably. Also, we use $U(\sigma^1; \sigma^2)$
to denote the payoff of a node when it follows strategy $\sigma^1$ while every other node
follows strategy $\sigma^2$. Note that a symmetric protocol yields the same payoff
to every node, thus achieving fairness among nodes.

\subsection{Deviation-Proof Protocols and the Efficiency Loss}

The goal of this paper is to build a protocol that fulfills the following
two requirements: (i) selfish nodes do not gain from manipulating the protocol, and (ii) the protocol achieves an optimal outcome.
We formalize the first requirement using the concept of deviation-proofness while evaluating
the second requirement using the concept of efficiency loss.
\begin{definition} \label{def:dp}
A protocol $\sigma \in \Sigma_i$ is \emph{deviation-proof} (DP) against a strategy $\sigma' \in \Sigma_i$ if
\beqn \nonumber
U(\sigma;\sigma) \geq U(\sigma';\sigma).
\eeqn
\end{definition}
When $\sigma$ is DP against $\sigma'$, a node cannot
gain by deviating to $\sigma'$ while other nodes follow $\sigma$.
Hence, if a deviating node has only one possible deviation strategy $\sigma'$,
a protocol $\sigma$ that is DP against $\sigma'$ satisfies the first requirement.
However, in principle, a deviating node can choose any strategy in $\Sigma_i$,
in which case we need a stronger concept than deviation-proofness.

Let $\Sigma_{c} \subset \Sigma_i$ be the set of all constant strategies that prescribe a fixed
transmission probability $p_d$, called the \emph{deviation probability}, in every slot
regardless of the history.
\begin{definition}
A protocol $\sigma \in \Sigma_i$ is \emph{robust $\epsilon$-deviation-proof} (robust $\epsilon$-DP) if
\beqn \nonumber
U(\sigma;\sigma) + \epsilon \geq U(\sigma';\sigma) \quad \text{for all $\sigma' \in \Sigma_c$}.
\eeqn
\end{definition}
In words, if a protocol $\sigma$ is robust $\epsilon$-DP, a node cannot gain more than $\epsilon$
by deviating to a constant strategy using a fixed deviation probability.
If there is a fixed cost of manipulating a given protocol and a deviation
strategy is constrained to constant strategies, then
a robust $\epsilon$-DP protocol can prevent a deviation by having $\epsilon$ smaller
than the cost. When there is no restriction on possible deviation strategies,
the following concept is relevant.
\begin{definition}
A protocol $\sigma \in \Sigma_i$ is \emph{$\epsilon$-Nash equilibrium} ($\epsilon$-NE) if
\beqn \nonumber
U(\sigma;\sigma) + \epsilon \geq U(\sigma';\sigma) \quad \text{for all $\sigma' \in \Sigma_i$}.
\eeqn
\end{definition}

We define the system payoff as the sum of the payoffs of all the
nodes in the system. Then, the system payoff when all nodes follow
a protocol $\sigma$ is given by $V(\sigma) \triangleq N U(\sigma; \sigma)$.
Since $N u^{\rm PO}$ is the maximum system payoff in the stage game achievable with a
symmetric action profile, we measure the efficiency loss from using a protocol by the following
concept.
\begin{definition}
The \emph{efficiency loss} of a protocol $\sigma \in \Sigma_i$ is defined as
\beqn \label{priceofdeviation}
C(\sigma) = N u^{\rm PO} - V(\sigma).
\eeqn
\end{definition}
\begin{definition}
A protocol $\sigma \in \Sigma_i$ is \emph{$\delta$-Pareto optimal} ($\delta$-PO) if
\beqn \nonumber
C(\sigma) \leq \delta.
\eeqn
\end{definition}

\begin{table*}
\caption{Main Results} 
\centering 
\vline
\begin{tabular}{c||c|c|c|c|}
\hline
Section & Signal & Test & Robustness to selfish manipulation & Optimality\\
\hline \hline
\multirow{2}{*}{III (Proposition 2)} & \multirow{2}{*}{Private (general)} & \multirow{2}{*}{Asymptotically perfect test} & DP against a strategy using & \multirow{2}{*}{$\delta$-PO} \\
& & & a constant transmission probability & \\
\hline
IV (Theorem 2) & Private (ACK feedback) & ACK ratio test & robust $\epsilon$-DP & $\delta$-PO \\
\hline
\multirow{2}{*}{V (Proposition 5)} & \multirow{2}{*}{Public (general)} & \multirow{2}{*}{Asymptotically perfect test} & DP against a strategy using a constant & \multirow{2}{*}{$\delta$-PO} \\
& & & transmission probability in a review phase & \\
\hline
VI (Theorem 4) & Public (ternary feedback) & Idle slot ratio test & $\epsilon$-NE & $\delta$-PO \\
\hline
\end{tabular} \label{table:summary} 
\end{table*}

A $\delta$-PO protocol is a protocol that yields an efficiency loss less than
or equal to $\delta$.
Let $\sigma^c$ be the strategy that prescribes the cooperation probability $p_c$
in every slot regardless of the history.\footnote{Note that $\sigma^c$ corresponds
to a slotted Aloha protocol that does not distinguish new and backlogged packets as in \cite{Jin1}.}
Then $U(\sigma^c;\sigma^c)=u^{\rm PO}$, and thus $\sigma^c$ achieves full
efficiency (i.e., $0$-PO). However, $\sigma^c$ is not DP against a constant deviation strategy
with $p_d > p_c$ as a deviating node can increase its payoff from $p_c (1-p_c)^{N-1}$ to $p_d (1-p_c)^{N-1}$.
We construct DP protocols that achieve a near-optimal system payoff in the following sections,
whose main results are summarized in Table~\ref{table:summary}.

\section{Deviation-Proof Protocols When Signals are Private}

\subsection{Description of Protocols with Private Signals} \label{sec:private}

In this section, we consider private signals. As pointed out in \cite{Kandori}, when signals are private,
it is difficult, if not impossible, to construct a NE that has a simple structure and
is easy to compute. Thus, we focus on a simpler problem of constructing a DP protocol
against a constant deviation strategy $\sigma^d \in \Sigma_c$.
Since a simple protocol such as $\sigma^c$ is DP against $\sigma^d$ with
$p_d \in [0,p_c]$, we restrict our attention to deviation strategies with $p_d \in (p_c,1]$.
Note that the restriction to constant deviation strategies is relevant when a deviating node
has a limited deviation capability in the sense that it can reset its transmission
probability only at the beginning.

\begin{figure}
\begin{center}
\includegraphics[width=4cm]{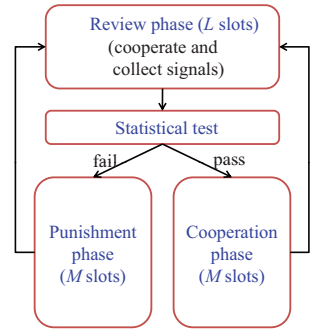}
\caption{Review strategy with private signals} \label{FigRev}
\end{center}
\vspace{-0.35cm}
\end{figure}

We build a protocol based on a review strategy.
When a node uses a review strategy, it starts from a review phase for which
it transmits with probability $p_c$ and collects signals. When the review phase ends, the node performs a statistical
test whose null hypothesis is that every node transmitted with probability $p_c$
during the review phase, using the collected signals. Then, the node moves to a reciprocation phase for which
it transmits with probability $p_c$ (cooperation phase) if the test is passed and with probability 1
(punishment phase) if the test fails. When the reciprocation phase ends, a new review phase begins.
A review strategy, denoted by $\sigma^r$, can be characterized by three elements, $(R,L,M)$,
where $R$ is a statistical test, and $L$ and $M$ are natural numbers that represent the lengths of
a review phase and a reciprocation phase, respectively. Thus, we sometimes write $\sigma^r$ as
$\sigma^{r}(R,L,M)$. With a protocol based on review strategy $\sigma^{r}(R,L,M)$, each node
performs the statistical test $R$ after slot $l(L+M)+L$ based on the signals $(z_i^{l(L+M)+1},\ldots,z_i^{l(L+M)+L})$ collected
in the recent review phase, for $l=0,1,\ldots$. A schematic representation of a review strategy with private signals is provided in Fig.~\ref{FigRev}.

The review strategies in \cite{RoyRadner} differ
from the review strategies described above in that in \cite{RoyRadner} a new review phase begins without having
a reciprocation phase if the test is passed. A key difference between the model of \cite{RoyRadner}
and ours is that in the principal-agent model of \cite{RoyRadner} only the principal reviews the performance
of the agent whereas in our model multiple nodes simultaneously review the performance of other nodes.
When signals are private, nodes do not know the results of the test
performed by other nodes. Hence, without a reciprocation phase followed by a successful
review, nodes cannot distinguish a deviating node from a punishing node and thus
cannot coordinate to begin a new review phase. This problem can be avoided when
signals are public, because the results of the test are the same across nodes in the case of public signals.\footnote{Alternatively,
this problem can be avoided by having a node that has a failed test broadcast that it moves
to a punishment phase, as in \cite{RoyRadner1}. However, this requires communication among
nodes, which we do not allow in this paper.}
Thus, a review strategy is modified accordingly in Section V, where we consider public signals.

\subsection{Analysis of Protocols with Private Signals}

\subsubsection{Existence of Deviation-Proof Protocols}

For the sake of analysis, we consider a fixed constant deviation strategy $\sigma^d \in \Sigma_c$
and the corresponding deviation probability $p_d \in (p_c, 1]$.
Given a symmetric protocol that prescribes a review strategy, we can compute
two probabilities of errors.
\begin{itemize}
\item False punishment probability $P_f(R,L)$: probability that
there is at least one node whose test fails after a review phase
when nodes follow a protocol $\sigma^r$.
\item Miss detection probability $P_m(R,L; p_d)$: probability that
there is no node among those following $\sigma^{r}$ whose test fails after a review phase
when there is exactly one node deviating to $\sigma^d$.
\end{itemize}
Since the payoff of every node is zero when there are two or more
punishing nodes, we need to have a small false punishment probability
to achieve a small efficiency loss.
On the other hand, in order to punish a deviating node effectively,
we need to have a small miss detection probability. Indeed, as will be shown
in Proposition 2, achieving small $P_f$ and $P_m$ is sufficient to design a
near-optimal DP protocol.

The payoff of a node when every node follows a review strategy $\sigma^r$ is given by
\begin{align*} 
U(\sigma^{r}; \sigma^{r}) =& \frac{(1-p_c)^{N-1}}{L+M} \biggl(p_c L +  p_c \left(1-P_{f} \right) M \nonumber \\
&+ \left( (1-P_f)^{\frac{N-1}{N}} \left( 1- (1-P_f)^{\frac{1}{N}} \right) \right) M \biggr).
\end{align*}
The payoff of a node choosing deviation strategy $\sigma^d$ while other nodes follow $\sigma^r$ is given by
\beqn 
U(\sigma^d; \sigma^{r}) = \frac{p_d(1-p_c)^{N-1}}{L+M} \left(L + P_m M \right). \nonumber
\eeqn
By Definition~\ref{def:dp}, $\sigma^r$ is DP against $\sigma^d$ if and only if
\beqn \label{LM}
U(\sigma^{r}; \sigma^{r}) \geq U(\sigma^d; \sigma^{r}).
\eeqn
The following theorem provides a necessary and sufficient condition
for a review strategy to be DP against $\sigma^d$.

\begin{theorem}
Given $p_d \in (p_c, 1]$, protocol $\sigma^r(R,L,M)$ is DP against
$\sigma^d$ if and only if $g(R,L;p_d) > 0$ and $M \geq M_{\rm min}(R,L;p_d)$,
where
\begin{align}
g(R,L;p_d) \triangleq & \left(1-P_{f}(R,L)\right)^{\frac{N-1}{N}} - (1-p_c) \left(1-P_f(R,L)\right)  \nonumber \\
& - p_d P_m(R,L;p_d) \label{pun3}
\end{align}
and
\beqn
M_{\rm min}(R,L;p_d) \triangleq \frac{(p_d-p_c)L}{g(R,L;p_d)}. \nonumber 
\eeqn
\end{theorem}

\begin{IEEEproof}
Note that the net payoff gain from deviating to the deviation strategy $\sigma^d$ is given by
\begin{align}
&U(\sigma^d; \sigma^{r}) - U(\sigma^{r}; \sigma^{r}) \nonumber \\ =& \;\;\frac{(1-p_c)^{N-1}}{L+M}\Bigl((p_d - p_c)L - g(R,L;p_d) M \Bigr). \label{eq:gain}
\end{align}
The first term in \eqref{eq:gain} is the gain during a review phase while the second
term is the loss during a reciprocation phase.
By \eqref{LM}, $\sigma^r$ is DP against $\sigma^d$ if and only if $(p_d - p_c)L \leq g(R,L;p_d) M$.
It is easy to check that $g(R,L;p_d) > 0$ and $M \geq M_{\rm min}(R,L;p_d)$ imply $(p_d - p_c)L \leq g(R,L;p_d) M$.
Suppose that $(p_d - p_c)L \leq g(R,L;p_d) M$. Since $(p_d - p_c)L > 0$, we must have $g(R,L;p_d) > 0$, which in turn implies $M \geq M_{\rm min}(R,L;p_d)$.
\end{IEEEproof}

Theorem 1 shows that for a given statistical test $R$, we can construct a DP protocol
based on the test if and only if there exists a natural number $L$ such that $g(R,L;p_d) > 0$. Once we find
such $L$, we can use it as the length of a review phase and then choose a natural number $M$ satisfying
$M \geq M_{\rm min}(R,L;p_d)$ to determine the length of a reciprocation phase. An immediate
consequence of Theorem 1 is that if protocol $\sigma^r(R,L,M)$ is DP against $\sigma^d$, then
protocol $\sigma^r(R,L,M')$ with $M' \geq M$ is also DP against $\sigma^d$.
Thus, $M_{\rm min}(R,L;p_d)$ can be interpreted as the minimum length of a reciprocation
phase to make $\sigma^r(R,L,M)$ DP against $\sigma^d$.
The following result provides a sufficient condition on $R$ under which we can find $L$
such that $g(R,L;p_d) > 0$ and thus a DP protocol based on $R$ can be constructed.

\begin{corollary} Given $p_d \in (p_c, 1]$, suppose that $R$ satisfies $\lim_{L \rightarrow \infty}
P_f(R,L) = 0$ and $\lim_{L \rightarrow \infty} P_m(R,L;p_d) = 0$.
Then there exists $L$ such that $g(R,L;p_d) > 0$.
\end{corollary}

\begin{IEEEproof}
By \eqref{pun3}, $\lim_{L \rightarrow \infty}
P_f(R,L) = 0$ and $\lim_{L \rightarrow \infty} P_m(R,L;p_d) = 0$ imply that
$\lim_{L \rightarrow \infty} g(R,L;p_d) = p_c > 0$. Thus, $g(R,L;p_d) > 0$
for sufficiently large $L$.
\end{IEEEproof}

Combining Theorem 1 and Corollary 1, we can see that if test $R$ is ``asymptotically perfect'' in the sense that
the two probabilities of errors converge to zero as the test is performed using more
signals, then we can always design a review strategy based on $R$ that is
DP against $\sigma^d$.

\subsubsection{Near-Optimal Deviation-Proof Protocols}

Suppose that every node follows a review strategy $\sigma^r$.
Since signals provide only imperfect information about the transmission
probabilities of other nodes, it is possible that a punishment is triggered,
which results in an efficiency loss as confirmed in the following proposition.
We use $\Sigma_r$ to denote the set of
all review strategies with private signals.

\begin{proposition}
$C(\sigma^r) \geq 0$ for all $\sigma^r \in \Sigma_r$ (with equality if and only if $P_f = 0$).
\end{proposition}

\begin{IEEEproof}
Fix a protocol $\sigma^r(R,L,M) \in \Sigma_r$. By \eqref{priceofdeviation},
we can express the efficiency loss of $\sigma^r$ as
\begin{align}
C(\sigma^r) =& \frac{NM}{L+M} (1-p_c)^{N-1} \nonumber \\
& \Bigl( p_cP_{f} - (1-P_f)^{\frac{N-1}{N}} + (1 -P_f) \Bigr).\label{eq:effloss}
\end{align}
Since $(1-P_f)^{\frac{N-1}{N}}$ is concave, we have $(1-P_f)^{\frac{N-1}{N}} \leq
1 - \frac{N-1}{N} P_f$ for $P_f \in [0,1]$, with equality if and only if $P_f = 0$.
Using $p_c = 1/N$, we obtain the result.
\end{IEEEproof}

Proposition 1 says that there is always a positive efficiency loss resulting from
a review strategy
unless there is a perfect statistical test in the sense that punishment
is never triggered when every node follows $\sigma^r$ (i.e., $P_f = 0$).
Punishment results in an efficiency loss because the system payoff is the same
as $N u^{\rm PO}$ when there is only one punishing node while it is zero when there are
two or more. Hence, a longer punishment induces a larger efficiency loss.
As can be seen from \eqref{eq:effloss}, for given $R$ and $L$, $C(\sigma^r)$ is non-decreasing
(and increasing if $P_f > 0$) in $M$.
Therefore, if we find $(R,L)$ such that $g(R,L;p_d)>0$, choosing $M = \lceil M_{\rm min}(R,L;p_d) \rceil$
minimizes the efficiency loss while having $\sigma^r(R,L,M)$ DP against
$\sigma^d$, where $\lceil \cdot \rceil$ denotes the ceiling
function.
This observation
allows us to reduce the design choice from $(R,L,M)$ to $(R,L)$. The following
proposition provides a sufficient condition on the statistical test for
constructing a near-optimal DP protocol.

\begin{proposition}
Given $p_d \in (p_c, 1]$, suppose that $R$ satisfies $\lim_{L \rightarrow \infty}
P_f(R,L) = 0$ and $\lim_{L \rightarrow \infty} P_m(R,L;p_d) = 0$. Then for any
$\delta > 0$, there exist $L$ and $M$ such that $\sigma^r(R,L,M)$ is
DP against $\sigma^d$ and $\delta$-PO.
\end{proposition}

\begin{IEEEproof}
Since $\lim_{L \rightarrow \infty} g(R,L;p_d) = p_c > 0$, there exists $L_1$ such that $g(R,L;p_d) > 0$ for
all $L \geq L_1$. By Theorem 1, $\sigma^r(R,L,\lceil M_{\rm min}(R,L;p_d) \rceil)$ is DP against $\sigma^d$
for all $L \geq L_1$. Since $C(\sigma^r)$
is non-decreasing in $M$, we have
\begin{align}
0 \leq C(\sigma^r) \leq & \ \frac{N \left(M_{\rm min}(R,L;p_d) + 1\right)}{L+\left(M_{\rm min}(R,L;p_d) + 1\right)} (1-p_c)^{N-1} \nonumber \\
&\left[p_cP_{f}- (1-P_f)^{\frac{N-1}{N}} + (1 -P_f) \right]. \label{eq:loss}
\end{align}
Note that $\lim_{L \rightarrow \infty} M_{\rm min}(R,L;p_d)/L = (p_d - p_c)/p_c$, and thus
the right-hand side of \eqref{eq:loss} converges to zero as $L$ goes to infinity,
which implies $\lim_{L \rightarrow \infty} C(\sigma^r) = 0$.
Therefore, there exists $L_2$ such that $C(\sigma^r) < \delta$
for all $L \geq L_2$. Choose $L \geq \max \{L_1, L_2\}$ and
$M = \lceil M_{\rm min}(R,L;p_d) \rceil$ to obtain a protocol with the desired properties.
\end{IEEEproof}

Proposition 2 shows that the efficiency loss of a DP protocol can be made
arbitrarily small when there is an asymptotically perfect statistical test.
It also points out a trade-off between optimality and implementation cost. In order
to make the efficiency loss within a small desired level, $L$ should be chosen sufficiently
large, which requires large $M$ by the relationship $M = \lceil M_{\rm min}(R,L;p_d) \rceil$.
At the same time, as $L$ and $M$ become larger, each node needs to maintain
longer memory to execute a review strategy, which can be considered
as higher implementation cost.

We make a couple of remarks. First,
the constructed DP protocols are DP against multiple nodes deviating to $\sigma^d$.
The payoff gain from deviation decreases with the number of deviating nodes.
Hence, if a protocol can deter a single node from deviating to $\sigma^d$, it
can also deter multiple nodes from doing so. Second, the constructed DP protocols are
DP against a more general class of deviation strategies with which a permanent deviation
to $p_d$ occurs in an arbitrary slot (determined deterministically or randomly).
A deviating node cannot gain starting from a review phase after a deviation occurs,
and without discounting its temporary gain is smaller than the perpetual loss.

\section{Protocols Based on the ACK Ratio Test}

\subsection{Description of the ACK Signal Structure and Protocols Based on the ACK Ratio Test}

In this section, we illustrate the results in Section III by considering
a particular signal structure and a particular statistical test.
In the slotted Aloha protocol in \cite{Roberts},
a node receives an acknowledgement (ACK) signal if it transmits its packet successfully and no signal otherwise. In the ACK signal
structure, the signal space can be written as $Z_i = \{S,F\}$,
for all $i \in \mathcal{N}$, where $z_i = S$ means that node $i$ receives an ACK signal and
$F$ means that it does not. We assume that there is no error in the transmission and reception of
ACK signals. The signal distribution $Q$ is such that $Q(\mathbf{a})$ puts probability mass
1 on $\mathbf{z} \in \mathcal{Z}$ with $z_i = S$ and $z_j = F$ for all $j \neq i$ if $a_i = T$ and $a_j = W$ for all $j \neq i$,
for each $i \in \mathcal{N}$, and probability mass 1 on $z_i = F$ for all $i$ otherwise.
Since it is possible for nodes to receive different signals (only one node receives signal $S$
when a success occurs), ACK signals are private.

\begin{figure}
\begin{center}
\vspace{4cm}
\includegraphics[width=8.9cm, height=7cm]{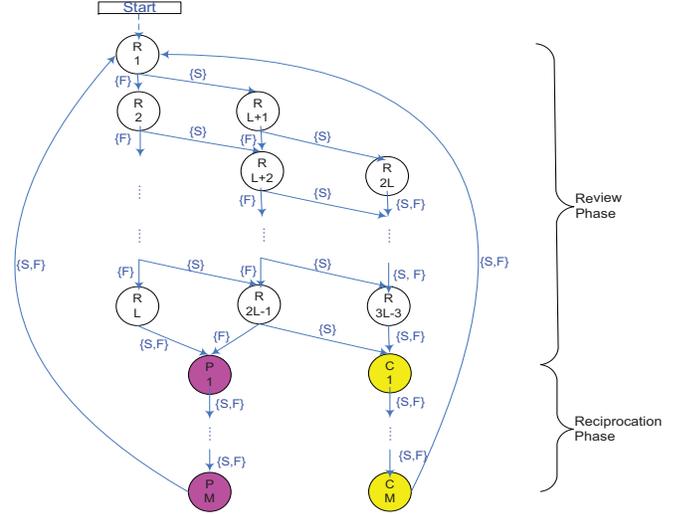}
\vspace{-3.8cm}
\caption{Automaton representation of a review strategy based on the ACK ratio test with parameters satisfying $1 \leq L(q_c - B) < 2$.} \label{Fig0}
\end{center}
\vspace{-0.35cm}
\end{figure}

In a review strategy with the ACK signal structure, a node uses its
ACK signals collected in a review phase to perform a statistical test.
We propose a particular statistical test called the
ACK ratio test. The test statistic of the ACK ratio test is the ratio of the number
of ACK signals obtained in a review phase to the length of a review phase, i.e.,
$\sum_{k=1}^{L} \chi{\{z_i^{\tau+k} = S\}}/L$, where $\chi$ is an
indicator function and $\tau+1$ represents a slot when a
review phase begins. The test is passed if the
statistic exceeds a threshold value, $q_c - B$, where $q_c \triangleq
p_c(1-p_c)^{N-1}$ and $B \in (0,q_c)$, and fails otherwise.
Note that $q_c$ is the expected value of the ACK ratio when every node
transmits with probability $p_c$. If there is a deviating node, the
ACK ratio tends to be smaller because its expected value is reduced
to $q_d \triangleq p_c(1-p_c)^{N-2}(1-p_d)$. The ACK ratio
test is designed to distinguish between these two events statistically while having
$B$ as a ``margin of error.'' Since the ACK ratio test can be
identified with $B$, we use $B$ instead of $R$ to
represent the ACK ratio test.

A review strategy based on the ACK ratio test, $\sigma^{r}(B,L,M)$, can be represented formally as follows:
\beqa
\sigma^{r}(h^t_i) =  \left \{
\begin{array}{ll}
p_c, & \!\!\!\textrm{ $t \in [l(L+M)+1, l(L+M)+L]$,} \nonumber \\
1, & \textrm{$t \in [l(L+M)+L+1,(l+1)(L+M)]$}, \nonumber \\
&\quad  \sum_{k=l(L+M)+1}^{l(L+M)+L} \chi{\{z_i^k = S\}}/L \leq q_c - B \nonumber\\
p_c, & \textrm{$t \in [l(L+M)+L+1,(l+1)(L+M)]$}, \nonumber \\
&\quad  \sum_{k=l(L+M)+1}^{l(L+M)+L} \chi{\{z_i^k = S\}}/L > q_c - B \nonumber
\end{array} \right.
\eeqa
for $l = 0, 1, \ldots$. Fig.~\ref{Fig0} shows an automaton representation of the review strategy $\sigma^r$ for $1 \leq L(q_c - B) < 2$
so that a node triggers punishment if it obtains less than two successes in a review phase.
Each state transition is labeled by the set of signals
that induce the transition.
In a reciprocation phase, a node goes through either states P$1$ to P$M$ (punishment phase) or states C$1$ to C$M$
(cooperation phase) depending on the number of ACK signals obtained in the review phase. Note that the number of states in the automaton representation of protocol $\sigma^r(B,L,M)$
is given by $N_s(\sigma^r) = kL - k(k-1)/2 + 2M$, where $k \geq 2$ is the natural number satisfying $k-2 \leq L(q_c-B) < k-1$.

\subsection{Analytical Results} \label{strategy}

Let $F(y;n,p)$ be the cumulative distribution function of a binomial
random variable with total number of trials $n$ and probability of
success $p$, i.e.,
\begin{align*}
F(y;n,p) = \sum_{m=0}^{\lfloor y \rfloor}\dbinom{n}{m}
p^m(1-p)^{n-m},
\end{align*}
where $\lfloor \cdot \rfloor$ denotes the floor function. Suppose
that every node transmits with probability $p_c$ in a review phase.
Then, the number of ACK signals that a node receives in the review
phase follows a binomial distribution with parameters $L$ and $q_c$.
Thus, the probability that a punishment is triggered by node $i$ is
given by
\begin{align*}
\Pr\left\{\sum_{k=1}^{L} \chi{\{z_i^{\tau+k} = S\}}/L \leq q_c - B
\right\} = F(L(q_c-B);L,q_c),
\end{align*}
and the false punishment probability is given by
\begin{align*} 
P_{f}(B,L) = 1 - \left[1 - F(L(q_c-B);L,q_c)\right]^N.
\end{align*}
Suppose that there is exactly one deviating node using $\sigma^d$, i.e., transmitting with
probability $p_d$ in a review phase. Then, the success probability in
the binomial distribution changes to $q_d$, and thus the miss
detection probability is given by
\begin{align*} 
P_{m}(B,L;p_d) = \left[1 - F(L(q_c-B);L,q_d)\right]^{N-1}.
\end{align*}
The monotonicity of $P_f$ and $P_m$ with respect to the test parameter $B$ is readily obtained.
\begin{proposition}
Given $p_d \in (p_c, 1]$ and $L$, $P_f(B,L)$ and $P_m(B, L;p_d)$ are non-increasing and non-decreasing in $B \in (0,q_c)$, respectively.
\end{proposition}

\begin{IEEEproof}
The proof is straightforward by noting that $F(L(q_c-B);L,q_c)$ and $F(L(q_c-B);L,q_d)$ are non-increasing in $B \in (0,q_c)$.
\end{IEEEproof}
As the margin of error is larger, it is more likely that the test is passed,
yielding a smaller false punishment probability and a larger miss detection
probability. The following lemma examines the asymptotic properties of $P_f$ and $P_m$
as $L$ becomes large.

\begin{lemma}
Given $p_d \in (p_c, 1]$, $\lim_{L \rightarrow
\infty} P_f(B,L) = 0$ for all $B \in (0,q_c)$,
$\lim_{L \rightarrow \infty} P_m(B,L;p_d) = 0$ for all $B
\in (0,q_c-q_d)$, and $\lim_{L \rightarrow \infty} P_m(B,L;p_d) = 1$ for all $B
\in (q_c-q_d,q_c)$.
\end{lemma}

\begin{IEEEproof}
Since $\chi{\{z_i^{\tau+k} = S\}}$, for $k = 1,\ldots,L$, can be considered as $L$ i.i.d.
random variables, we can apply the strong law of large numbers to the ACK ratio \cite{Billingsley}.
When every node transmits with probability $p_c$, the ACK ratio converges almost
surely to $q_c$ as $L$ goes to infinity, which implies that the false
punishment probability goes to zero for all $B > 0$. When
there is exactly one node transmitting with probability $p_d$, the
ACK ratio of a node transmitting with probability $p_c$ converges
almost surely to $q_d$ as $L$ goes to infinity. Hence, if
$q_d < q_c - B$ (resp. $q_d > q_c - B$), the miss detection probability goes to
zero (resp. one).
\end{IEEEproof}

Lemma 1 provides a sufficient condition on the ACK ratio test to apply Proposition 2.

\begin{proposition}
Suppose that $B \in (0,q_c-q_d)$. For any
$\delta > 0$, there exist $L$ and $M$ such that
$\sigma^r(B,L,M)$ is DP against $\sigma^d$ and
$\delta$-PO.
\end{proposition}

\begin{IEEEproof}
The proposition follows from Lemma 1 and Proposition 2.
\end{IEEEproof}

Proposition 4 states that for given $p_d \in (p_c,1]$, we can construct
a protocol $\sigma^r$ that is DP against $\sigma^d$ and achieves an
arbitrarily small PoS by setting $B$ such that $0 < B
< q_c-q_d = p_c(1-p_c)^{N-2}(p_d-p_c)$. Note that as $p_d$ is
larger, it is easier to detect a deviation, and thus we have a wider
range of $B$ that renders deviation-proofness and near-optimality.


So far we have considered a constant deviation strategy $\sigma^d$ prescribing a fixed deviation probability $p_d$
and designed a protocol that is DP against $\sigma^d$. However,
it is natural to regard $p_d$ as a choice by a deviating node, and thus in principle it can be
any probability. Now we allow the possibility that a deviating node can use
any constant deviation strategy, and we obtain the following result.

\begin{theorem}
For any $\epsilon > 0$ and $\delta > 0$, there exist $B$, $L$, and $M$
such that $\sigma^r(B,L,M)$ is robust $\epsilon$-DP and $\delta$-PO.
\end{theorem}

\begin{IEEEproof}
The proof is relegated to Appendix A.
\end{IEEEproof}

We can interpret $\epsilon$ and $\delta$ as performance requirements.
Requiring smaller $\epsilon$ makes protocols more robust while
requiring smaller $\delta$ results in a higher system payoff.
In addition to the trade-off between optimality and implementation cost already
mentioned following Proposition 2, we can identify a similar trade-off between
robustness and implementation cost in that smaller $\epsilon$ in general requires
larger $L$ and $M$ to construct a robust $\epsilon$-DP protocol.

\subsection{Numerical Results}

\begin{figure*}
\centering
\subfloat[]{\includegraphics[width=9cm, height=5cm]{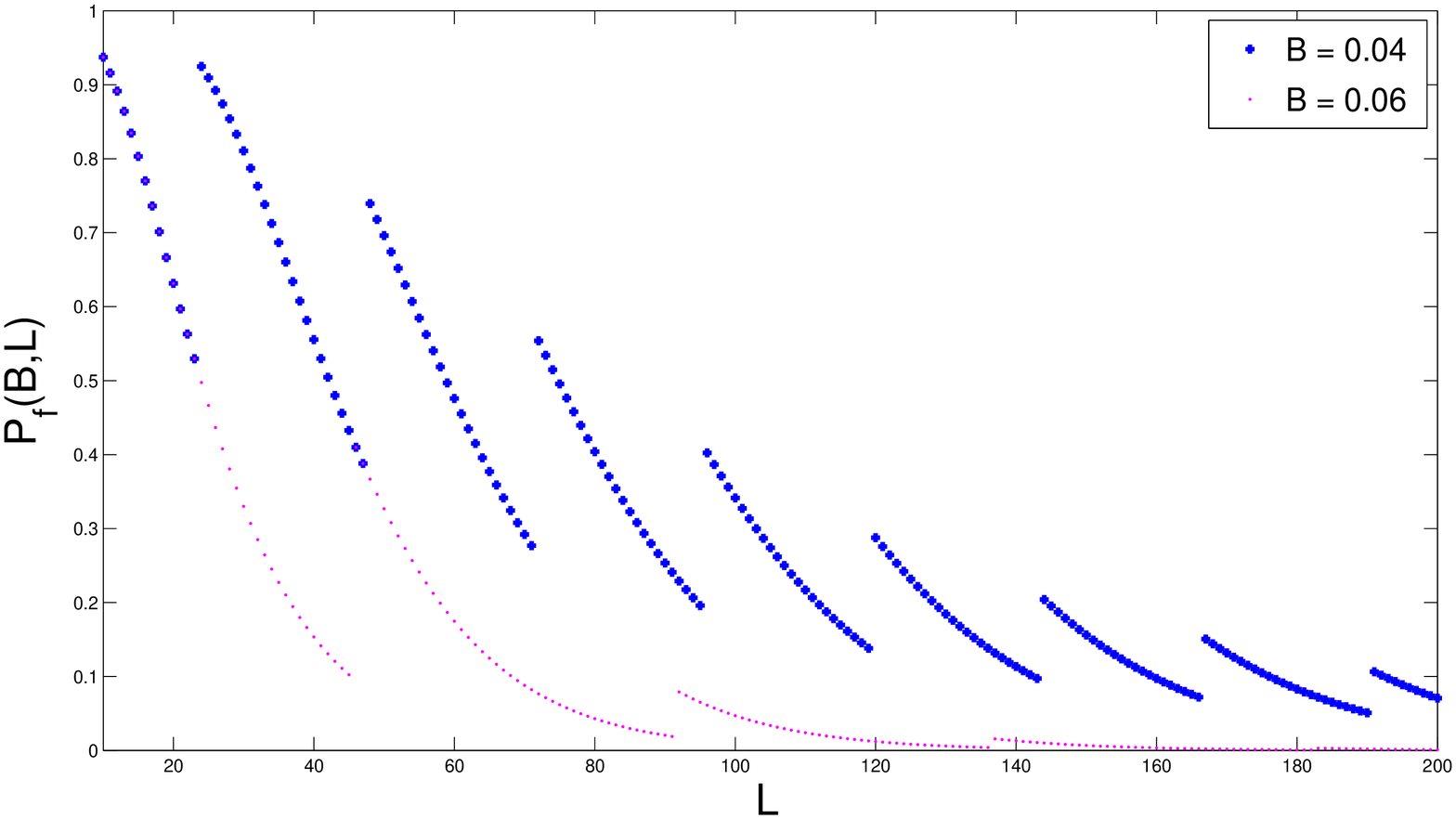}}
\subfloat[]{\includegraphics[width=9cm, height=5cm]{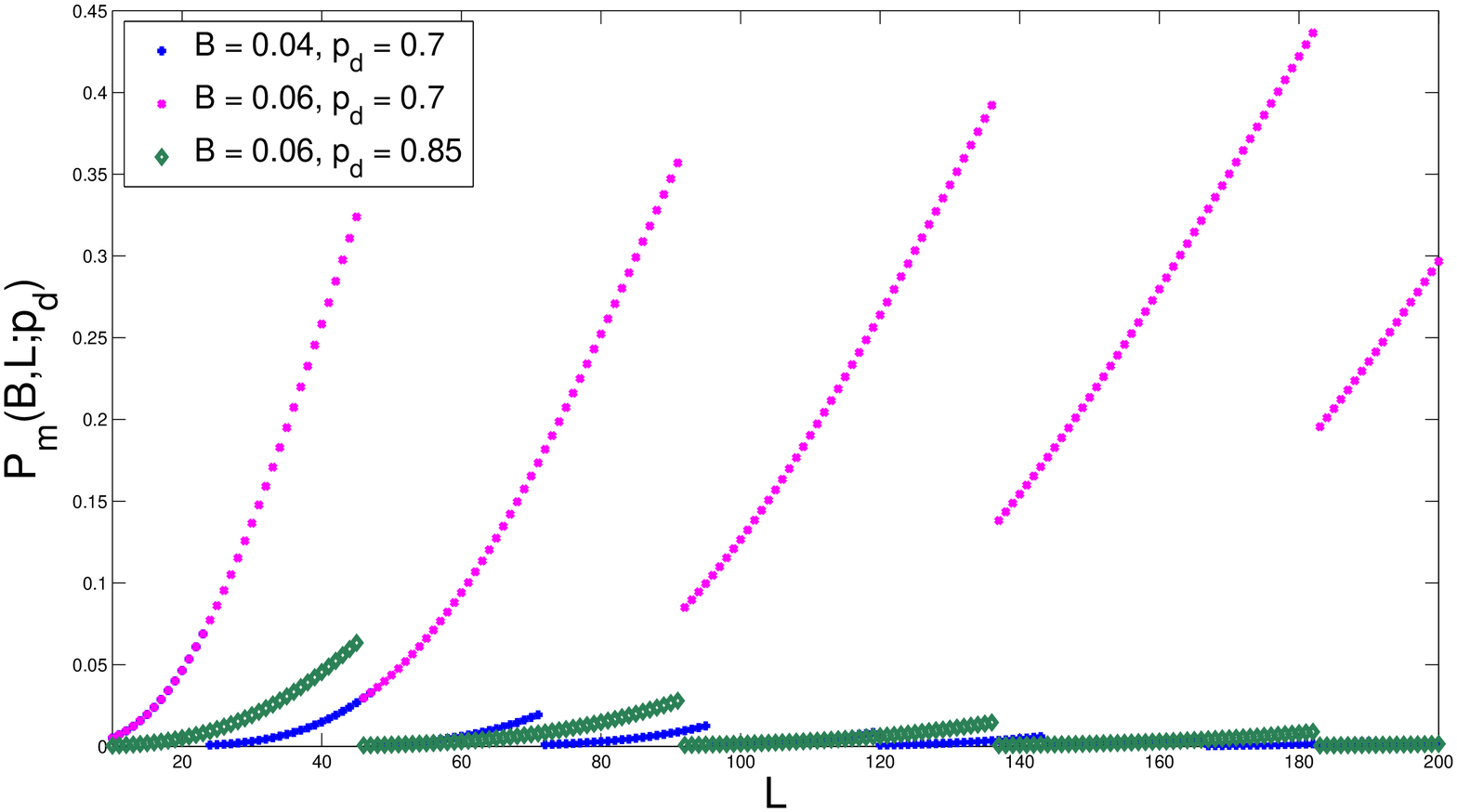}}
\caption{$P_f(B,L)$ and $P_m(B,L;p_d)$ versus the length of a review phase $L$.}
\label{Fig1}
\vspace{-0.35cm}
\end{figure*}

To provide numerical results, we consider a network with 5 nodes, i.e., $N=5$ and $p_c=1/N = 0.2$.
Fig.~\ref{Fig1} plots the false punishment probability $P_f(B,L)$ and the miss
detection probability $P_m(B,L;p_d)$ while varying the length of a review
phase $L$. Fig.~3(a) shows that $P_f(B,L)$ exhibits a
decreasing tendency as $L$ increases, with discontinuities occurring
at the points where the floor function of $L(q_c-B)$ has a
jump. We can also see that $P_f$ is smaller for larger $B$, as shown in Proposition 3.
The upper threshold for the parameter $B$ to yield $\lim_{L \rightarrow \infty} P_m(R,L;p_d) = 0$
in Lemma 1 is $q_c-q_d = 0.0512$ for $p_d = 0.7$.
We can see from Fig. 3(b) that $P_m(B,L;p_d)$ approaches 0 as $L$ becomes large
when $B$ is smaller than this threshold, whereas it approaches 1 when
$B$ exceeds the threshold.
Fig. 3(b) also shows that, for fixed $B$, $P_m(B,L;p_d)$ is smaller for larger
$p_d$, i.e., as the deviation becomes greedier, it
is more likely to be detected.

\begin{figure*}
\centering
\subfloat[]{\includegraphics[width=9cm, height=5cm]{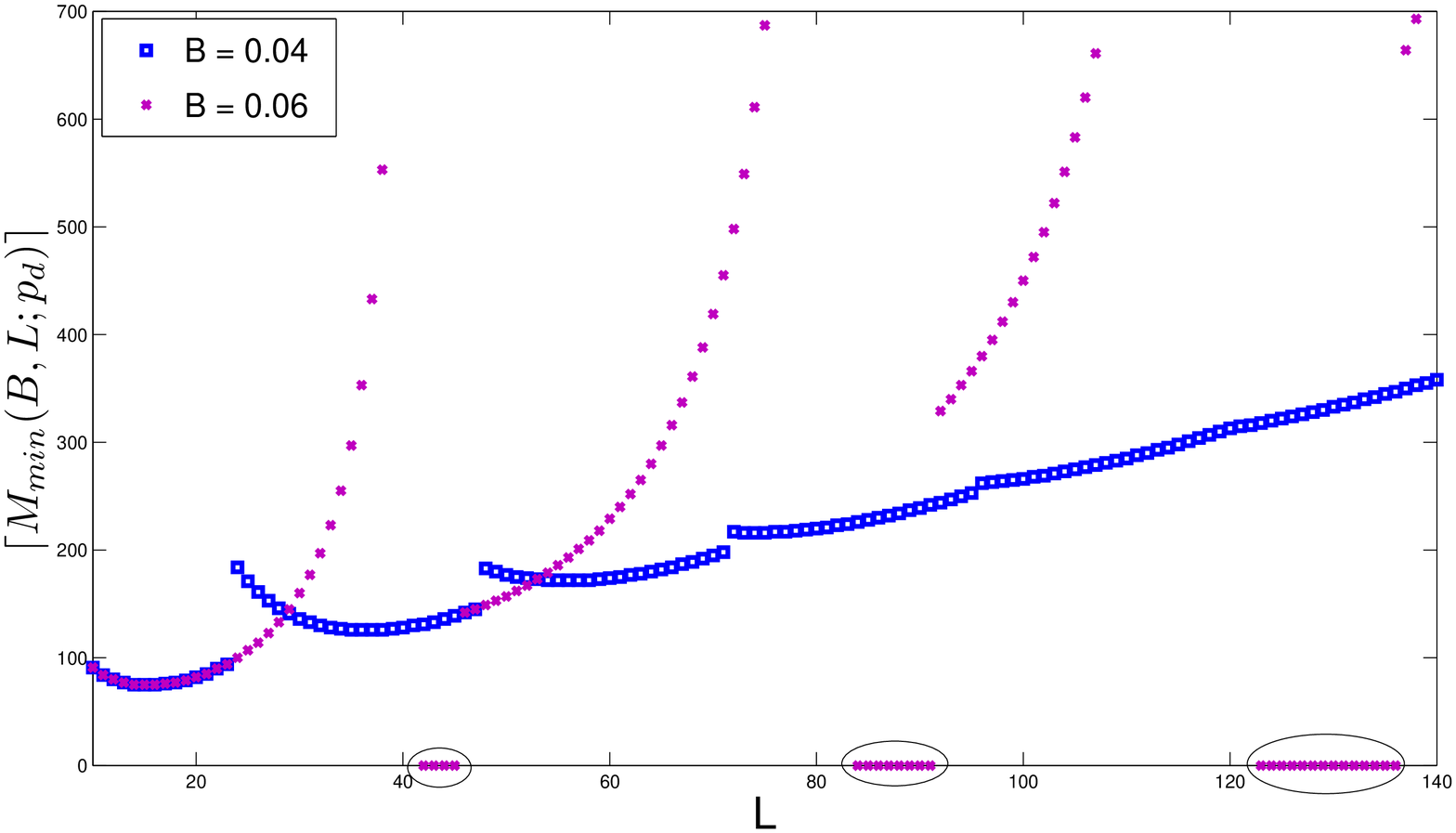}}
\subfloat[]{\includegraphics[width=9cm, height=5cm]{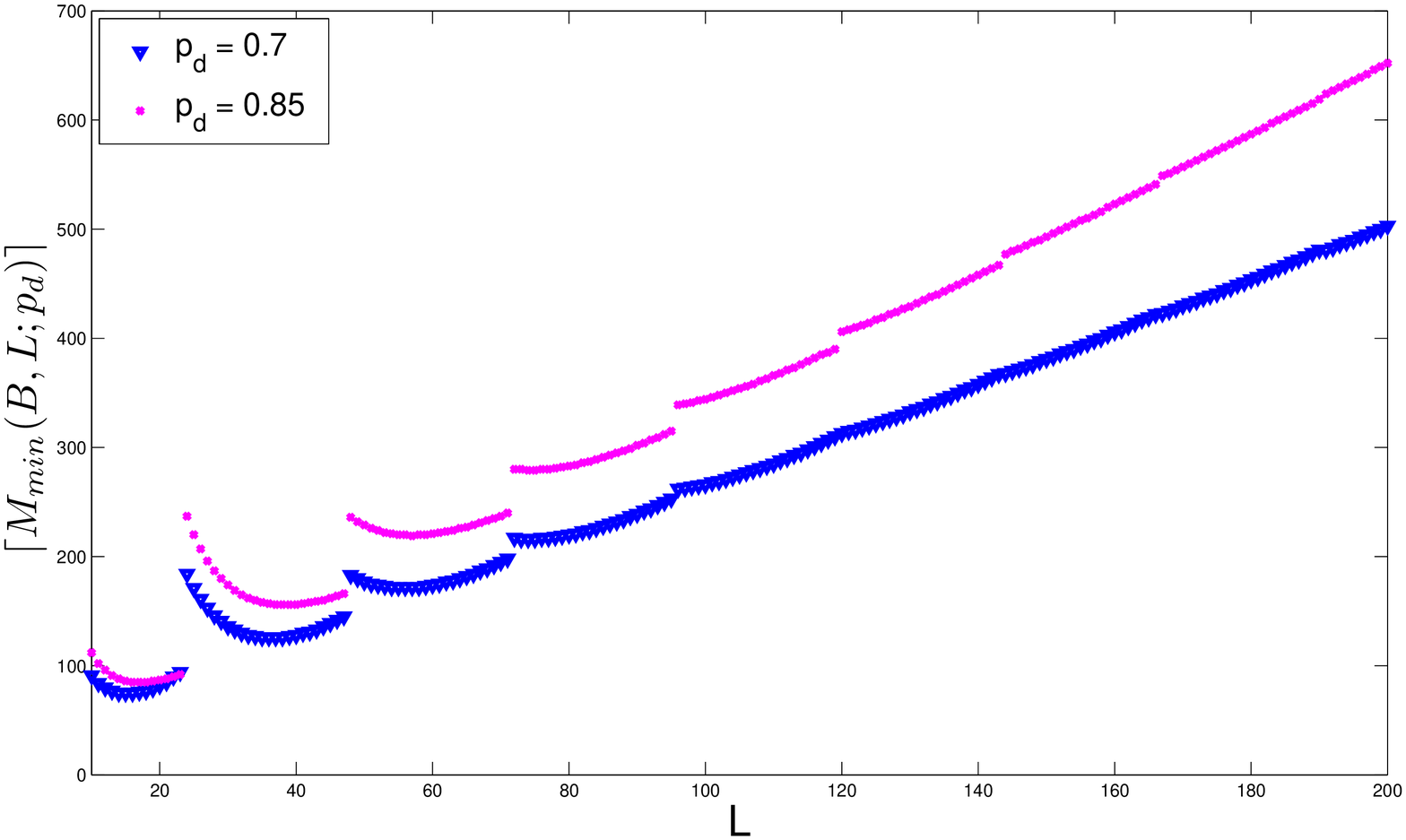}}
\caption{The minimum length of a reciprocation phase $\lceil M_{\rm min}(B,L;p_d) \rceil$ versus the length of a review phase $L$: (a) $p_d = 0.7$, and (b) $B = 0.04$.}
\label{Fig2}
\vspace{-0.35cm}
\end{figure*}

\begin{figure*}
\centering
\subfloat[]{\includegraphics[width=9cm, height=5cm]{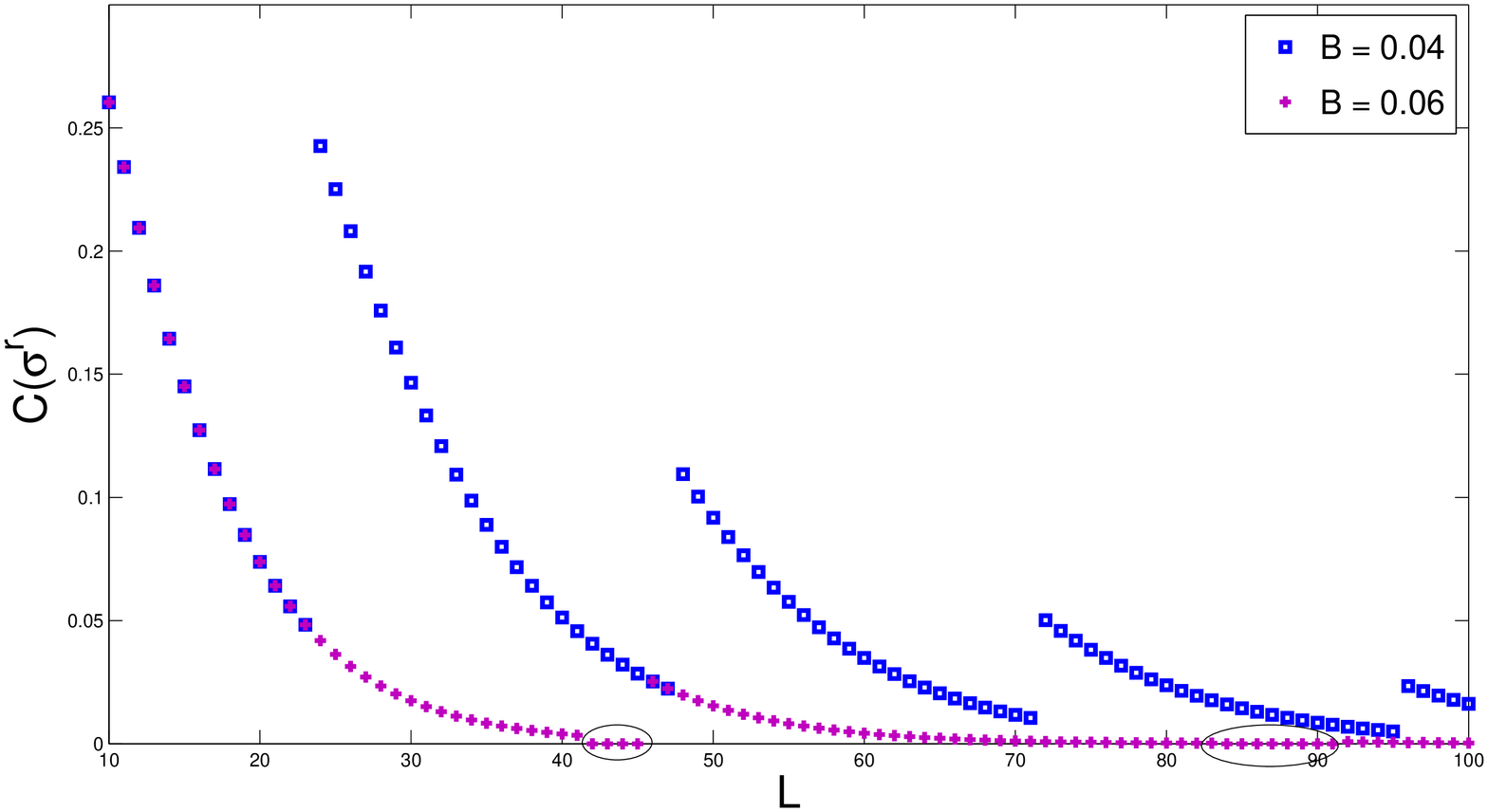}}
\subfloat[]{\includegraphics[width=9cm, height=5cm]{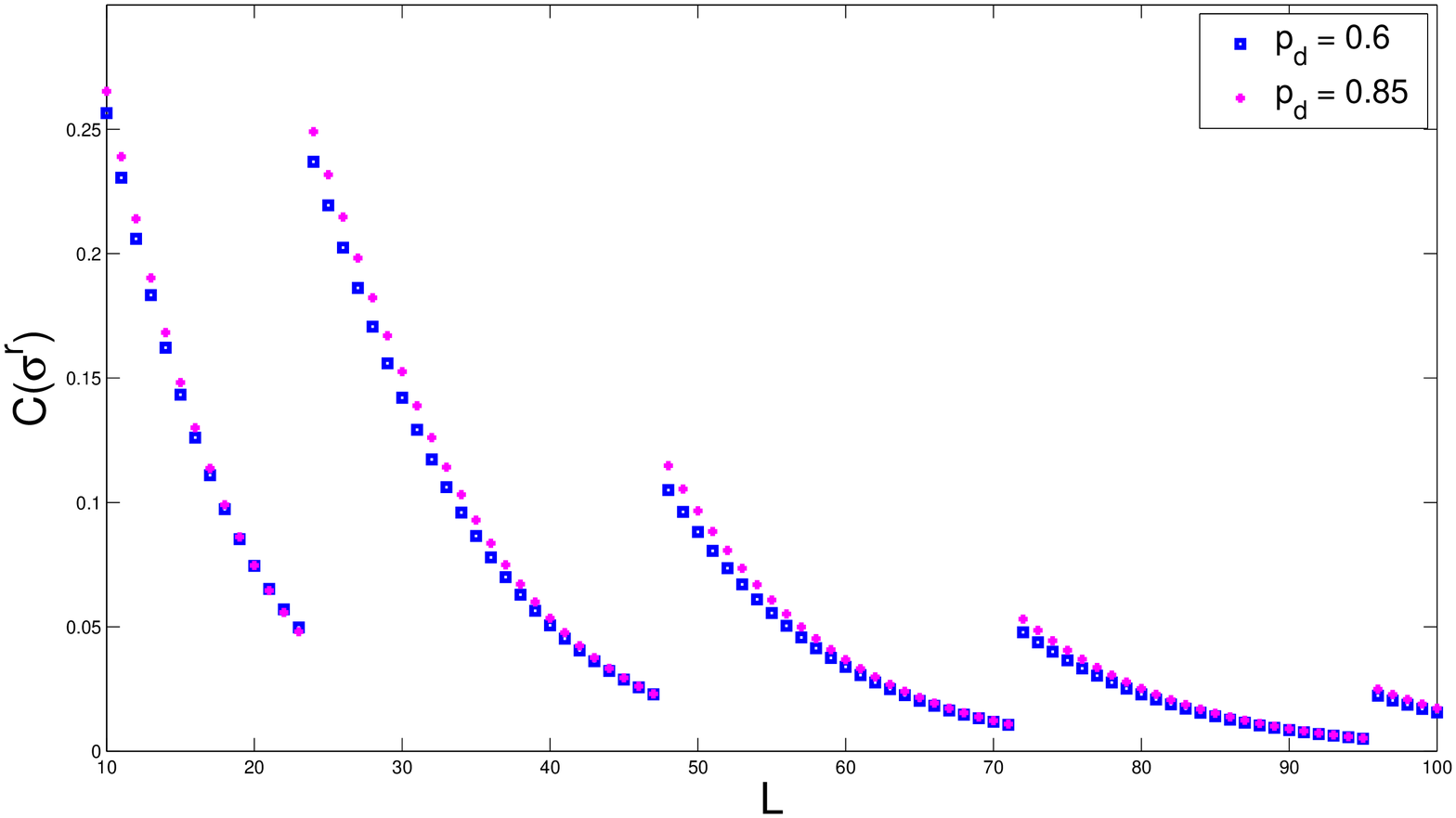}}
\caption{Efficiency loss $C(\sigma^r)$ versus the length of a review phase $L$: (a) $p_d = 0.7$, and (b) $B = 0.04$.}
\label{Fig3}
\vspace{-0.35cm}
\end{figure*}

Fig.~\ref{Fig2} plots the relationship between the length of a
review phase $L$ and the minimum length of a reciprocation phase
$\lceil M_{\rm min}(B,L;p_d) \rceil$ to have a DP protocol for different values of $B$ and $p_d$.
In Fig.~4(a), we fix $p_d = 0.7$ and consider $B = 0.04$ and $0.06$.
Note that when $B = 0.06$, some values of $L$ result in large minimum values
of $M$, which are not displayed in Fig.~4(a). Also, the values of $L$ with which
no DP protocol can be constructed for given $B$ and $p_d$ (i.e., $g(B,L;p_d) \leq 0$)
are indicated with $\lceil M_{\rm min}(B,L;p_d) \rceil=0$ in Fig.~4(a).
For example, we cannot construct a DP protocol using $L$ such that $42 \le L \le 45$ or
$84 \le L \le 91$ when $B = 0.06$ and $p_d = 0.7$. When $B=0.04$, we can construct a DP protocol using any $L \geq 10$.
In Fig. 4(b), we fix $B = 0.04$ and consider $p_d = 0.7$ and $0.85$. For the considered values of $p_d$,
we observe that the minimum length of a reciprocation phase is increasing in $p_d$ for the most values of $L$.
Also, in general, a longer review phase requires a longer reciprocation phase for fixed $p_d$
although a reverse relationship may be obtained, especially when $L$ is small.
Note that $L$ and $\lceil M_{\rm min}(B,L;p_d) \rceil$ have a linear relationship in the limit
since $\lim_{L \rightarrow \infty} M_{\rm min}(B,L;p_d)/L = (p_d - p_c)/p_c$.

\begin{table*}
\caption{Parameters and the Efficiency Loss of Optimal Protocols} 
\centering 
\vline
\begin{tabular}{l|c|c|c|c|c|c|c|c|c|}
\hline
$p_d$&0.6&0.65&0.7&0.75&0.8&0.85&0.9&0.95&1\\
\hline
$(L,M)$&(22,101)&(23,101)&(23,94)&(23,91)&(23,90)&(23,92)&(23,96)&(23,102)&(22,106)\\
\hline
$C(\sigma^r)$ &0.0570&0.0490&0.0483&0.0480&0.0479&0.0481&0.0485&0.0490&  0.0575\\
\hline
\end{tabular}\label{table:nonlin} 
\end{table*}

Fig. \ref{Fig3} plots efficiency loss $C(\sigma^r)$ against the
length of a review phase $L$ when the length of a reciprocation phase is
chosen as $\lceil M_{\rm min}(B,L;p_d) \rceil$ for different values of $B$ and $p_d$. The points
where efficiency loss is shown as 0 in Fig. 5(a) are where no DP
protocol exists for the given parameters. We can observe that as $L$
increases, efficiency loss tends to decrease to 0, which is consistent with
Proposition 4. Fig. 5(a) shows that for fixed $p_d = 0.7$, efficiency loss is smaller when $B = 0.06$ than when $B = 0.04$.
This is because the false punishment probability of the former case is smaller than that of the latter case as
shown in Fig. 3(a). Fig. 5(b) shows that efficiency loss is almost the same for the two considered deviation probabilities when $B = 0.04$.

\subsection{Deviation-Proof Protocols with Complexity Considerations} \label{design1}

\subsubsection{Protocol Design Problem with a Complexity Constraint} \label{design}

So far we have explored the possibility of constructing near-optimal
deviation-proof protocols based on a review strategy.
We mention briefly how to incorporate complexity considerations in the protocol
design problem. One approach to measure the complexity of a repeated game
strategy is to use the number of the states of the smallest automaton
that can implement the strategy \cite{Kalai}. Thus, we can formulate
the following protocol design problem, assuming that the deviation strategy
is fixed as $\sigma^d$.
\beqa
&&\!\!\!\!\!\!\!\!\!\!\!\!\text{minimize} \qquad \;\;C(\sigma^r(B,L,M)) \nonumber \\
\!\!\!&&\!\!\!\!\!\!\!\!\!\!\!\!\text{subject to} \qquad \;\;\;  \!\!\text{$\sigma^r$ is DP against $\sigma^d$} \label{optdesign} \\
&&  \quad \quad \quad \quad  \qquad \!\!\!\!\!\!\!\!\!N_s(\sigma^r) \le \overline{N}_s \nonumber
\eeqa
The second constraint can be interpreted as a complexity constraint which bounds
the number of states in the automaton representation of $\sigma^r$. Without a
complexity constraint, efficiency loss can be made arbitrarily small while
satisfying the first constraint by choosing sufficiently large $L$, as shown in Proposition 4.
Thus, the second constraint prevents $L$ from growing without bound.

\subsubsection{Protocol Design Method}

We propose a method to find an optimal protocol that solves the
protocol design problem \eqref{optdesign}.
\begin{itemize}
\item {\it Step 1}. Determine a finite set $\mathcal{B} \subset (0,q_c)$ as
the set of possible values of $B$.
\item {\it Step 2}. Fix $B \in \mathcal{B}$. Identify the set of feasible
$(L,M)$ in the sense that $(L,M)$ satisfies the second constraint of \eqref{optdesign}
given $B$.
\item {\it Step 3}. Fix feasible $L$, and check whether $g(B,L;p_d)$ in \eqref{pun3} is positive.
If so, choose $M$ as the smallest feasible value of $M$ larger than or equal to $M_{\rm min}(B,L;p_d)$, which we denote by
$M(B,L)$, if such a value exists. Then, $\sigma^r(B,L,M(B,L))$ is a protocol that satisfies both constraints of \eqref{optdesign}.
\item {\it Step 4}. By varying $B$ and $L$, obtain protocols that satisfy both constraints. Among these protocols, choose a protocol
that yields the smallest efficiency loss.
\end{itemize}
As an illustrative example, we consider $N=5$ and set $\overline{N}_s=2^{8}=256$ so that
protocols can be implemented using 8-bit memory. For simplicity, we fix $B$ at $0.04$, i.e., $\mathcal{B} = \{0.04\}$.
Table~\ref{table:nonlin} presents the parameters $(L, M)$ and the efficiency loss of optimal protocols for different
deviation probabilities.
We can see that the optimal protocols have different parameters for different values of $p_d$.
Due to jumps in the efficiency loss curves as shown in Fig.~\ref{Fig3}, the optimal protocols do not necessarily have the longest possible review phase.

\section{Deviation-Proof Protocols When Signals are Public}

\subsection{Motivation}

As mentioned in Section \ref{sec:private}, when signals are private,
nodes do not know the results of the test that other nodes perform. Hence, nodes need
to have a reciprocation phase regardless of the results of the test in order to
synchronize the beginning of a review phase across nodes. However, this structure of a review strategy
creates a weakness that can be exploited by a deviating node. A deviating node
can cooperate in a review phase to avoid punishment and then defect in a reciprocation
phase to obtain a payoff gain. To exclude such a ``smart'' deviation,
in Sections III and IV we have focused on constant deviation strategies when designing DP protocols.
However, this complication does not arise when signals
are public. Since the result of the test is commonly known among nodes,
a reciprocation phase can be skipped when the test is passed, eliminating the
room for exploitation. This added robustness of protocols with public signals
can be regarded as the \emph{value of public signals} when the signal structure
is a design choice.

\subsection{Description of Protocols with Public Signals}

When signals are public, nodes receive a common signal, and thus
we use $z^t$, without subscript $i$, to denote the signal in slot $t$.
A review strategy with public signals is the same as the one described in
Section \ref{sec:private} except that there is no cooperation phase. That is, a new review
phase begins immediately if the statistical test is passed. If the test fails,
a punishment phase occurs as before. Since we focus on symmetric protocols,
all nodes use the same statistical test and perform the test based on the same
signals. Hence, all nodes obtain the same result of the test, and thus they
are always in the same phase. We use $\tilde{\sigma}^{r}(R,L,M)$ to denote
the review strategy with public signals that uses test $R$ and has $L$ and $M$ as the lengths of
a review phase and a punishment phase, respectively.

\subsection{Analysis of Protocols with Public Signals}

We first consider a fixed deviation strategy $\tilde{\sigma}^d$ that has the same structure
as the prescribed review strategy $\tilde{\sigma}^{r}$. That is, a deviating node
transmits with probability $p_d$ in a review phase and with $p_r$ in a punishment phase.
Since no node obtains a positive payoff in a punishment phase,
the choice of $p_r$ does not affect the analysis, and thus
for analysis only $p_d$ matters.
For the same reason as in Section III, we focus on the case where $p_d > p_c$.

As in the case of private signals, we can compute
two probabilities of errors: the false punishment probability $\tilde{P}_f(R,L)$ and
the miss detection probability $\tilde{P}_m(R,L; p_d)$.
Since a punishment phase occurs with probability $\tilde{P}_f$ and results in zero payoff for every node
when all nodes follow a review strategy, we have
\beqn
U(\tilde{\sigma}^{r}; \tilde{\sigma}^{r}) = \frac{L q_c}{L+\tilde{P}_f M}. \nonumber
\eeqn
Note that $(L+\tilde{P}_f M)$ is the average length of an epoch, defined as a review phase
and the following punishment phase if one exists, and $L q_c$ is the accumulated
expected payoff for a node in an epoch.
The payoff of a node choosing deviation strategy $\tilde{\sigma}^d$ while other nodes
follow $\tilde{\sigma}^r$ is given by
\beqn 
U(\tilde{\sigma}^{d}; \tilde{\sigma}^{r}) = \frac{L q_d}{L+(1 - \tilde{P}_m) M}. \nonumber
\eeqn
The efficiency loss of $\tilde{\sigma}^{r}$ can be computed as
\begin{align} \label{eq:pos2}
C(\tilde{\sigma}^{r}) = \frac{N \tilde{P}_f M q_c}{L+\tilde{P}_f M},
\end{align}
which is always nonnegative (positive if $\tilde{P}_f > 0$). Note that the nonnegativity of
the efficiency loss does not require $p_c = 1/N$, unlike in the case of private signals (see Proposition 1).
The following theorem is an analogue of Theorem 1 for the case of public signals.

\begin{theorem}
Given $p_d \in (p_c, 1]$, protocol $\tilde{\sigma}^r(R,L,M)$ is DP against
$\tilde{\sigma}^d$ if and only if $\tilde{g}(R,L;p_d) > 0$ and $M \geq \tilde{M}_{\rm min}(R,L;p_d)$,
where
\begin{align} 
\tilde{g}(R,L;p_d) \triangleq p_c \bigl(1-\tilde{P}_m(R,L;p_d)\bigr) - p_d \tilde{P}_{f}(R,L) \nonumber
\end{align}
and
\beqn
\tilde M_{\rm min}(R,L;p_d) \triangleq \frac{(p_d-p_c)L}{\tilde{g}(R,L;p_d)}. \nonumber
\eeqn
\end{theorem}

\begin{IEEEproof}
$U(\tilde{\sigma}^{r}; \tilde{\sigma}^{r}) \geq U(\tilde{\sigma}^{d}; \tilde{\sigma}^{r})$
if and only if $(p_d-p_c)L \leq \tilde{g}(R,L;p_d) M$. Note that $(1-p_c)^{N-1}(p_d - p_c) L$
is the gain from deviation in a review phase while $(1-p_c)^{N-1} \tilde{g}(R,L;p_d) M$
is the expected loss from deviation in a punishment phase. The result can be obtained by
using a similar argument as in the proof of Theorem 1.
\end{IEEEproof}

Theorem 3 shows that for a given statistical test $R$, we can construct a DP protocol
based on the test if and only if there exists a natural number $L$ such that $\tilde{g}(R,L;p_d) > 0$.
Once we find such $L$, we can use it as the length of a review phase and then choose $M$ larger than
or equal to $\tilde{M}_{\rm min}(R,L;p_d)$ to determine the length of a punishment phase.
Since $C(\tilde{\sigma}^{r})$ is non-decreasing in $M$ for fixed $R$ and $L$
as can be seen from \eqref{eq:pos2},
the efficiency loss is minimized for given $(R,L)$ by setting $M = \lceil \tilde{M}_{\rm min}(R,L;p_d) \rceil$ so that the length of a punishment phase
is just enough to prevent deviation. Again, this observation reduces the design choices for
a review strategy from $(R,L,M)$ to $(R,L)$. The next result is an analogue of
Proposition 2, showing that if an asymptotically perfect statistical test is available,
we can construct a near-optimal DP protocol.

\begin{proposition}
Given $p_d \in (p_c, 1]$, suppose that $R$ satisfies $\lim_{L \rightarrow \infty}
\tilde{P}_f(R,L) = 0$ and $\lim_{L \rightarrow \infty} \tilde{P}_m(R,L;p_d) = 0$. Then for any
$\delta > 0$, there exist $L$ and $M$ such that $\tilde{\sigma}^r(R,L,M)$ is
DP against $\tilde{\sigma}^d$ and $\delta$-PO.
\end{proposition}

\begin{IEEEproof}
The proof is similar to that of Proposition 2, and thus is omitted for brevity.
\end{IEEEproof}

\section{Protocols Based on the Idle Slot Ratio Test}

\subsection{Description of the Ternary Signal Structure and Protocols Based on the Idle Slot Ratio Test}

\begin{figure}
\begin{center}
\vspace{4cm}
\includegraphics[width=8.9cm,height=7cm]{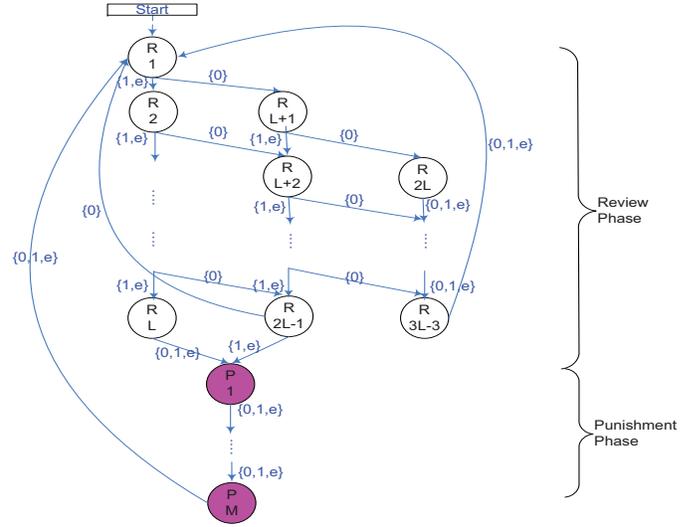}
\vspace{-3.8cm}
\caption{Automaton representation of a review strategy based on the idle slot ratio test with parameters satisfying $1 \leq L(\tilde q_c-B) < 2$.} \label{Fig00}
\end{center}
\vspace{-.35cm}
\end{figure}

To illustrate the results in Section V, we consider the ternary signal structure as in \cite{bert}, \cite{hajek},
whose signal space can be written as $Z = \{0,1,e\}$.
Nodes receive signal 0 if the slot is idle, 1 if there is a success, and $e$
if there is a collision. Signals under the ternary signal structure are public
because nodes always receive a common signal. We consider a review strategy
with which nodes use the fraction of idle slots in a review phase, or the idle slot ratio, as the
test statistics. If every node transmits with probability $p_c$, the expected
value of the idle slot ratio is $\tilde{q}_c \triangleq (1-p_c)^N$. On the other hand,
if there is exactly one deviating node that transmits with probability $p_d$
during a review phase, the expected value is reduced to
$\tilde{q}_d \triangleq (1-p_d)(1-p_c)^{N-1}$.
The idle slot ratio test is passed
if the idle slot ratio, $\sum_{k=1}^{L} \chi{\{z^{\tau+k} = 0\}}/L$, exceeds
a threshold value, $\tilde{q}_c - B$, and fails otherwise. Fig.~\ref{Fig00} shows an automaton
representation of a review strategy $\tilde \sigma^r$ whose parameters satisfying $1 \leq L(\tilde q_c-B) < 2$.
State transition occurs depending on the received signals, as depicted in Fig.~\ref{Fig00}.
When a review phase ends, nodes either start a new review phase or move to a punishment
phase depending on whether the number of idle slots in the review phase exceeds
$L(\tilde q_c-B)$ or not.

\subsection{Analytical Results}

Suppose that every node follows a review strategy based on the idle slot ratio
test, $\tilde{\sigma}^{r}(B,L,M)$.
Then, every node transmits with probability $p_c$ in a review phase, and the number of idle slots occurring in a review
phase follows a binomial distribution with parameters $L$ and $\tilde{q}_c$.
Thus, the false punishment probability is given by
\begin{align*}
\tilde{P}_{f}(B,L) = F(L(\tilde{q}_c-B);L,\tilde{q}_c).
\end{align*}
Since a deviating node using transmission probability $p_d$ changes
the ``success probability'' of the binomial distribution from $\tilde{q}_c$ to $\tilde{q}_d$, the miss detection probability is given by
\begin{align*}
\tilde{P}_{m}(B,L;p_d) = 1 - F(L(\tilde{q}_c-B);L,\tilde{q}_d).
\end{align*}
The monotonicity of $\tilde P_f$ and $\tilde P_m$ with respect to the margin of error $B$ is stated as follows.
\begin{proposition} Given $p_d \in (p_c, 1]$ and $L$, $\tilde P_f(B,L)$ and $\tilde P_m(B,L;p_d)$ are non-increasing and non-decreasing in $B \in (0,\tilde q_c)$, respectively.
\end{proposition}

\begin{IEEEproof} The proof is straightforward by noting that $F(L(\tilde{q}_c-B);L,\tilde{q}_c)$ and $F(L(\tilde{q}_c-B);L,\tilde{q}_d)$ is non-increasing in $B \in (0,\tilde q_c)$. \end{IEEEproof}

The next lemma examines the asymptotic properties of $\tilde{P}_f$ and $\tilde{P}_m$
as $L$ becomes large.

\begin{lemma}
Given $p_d \in (p_c, 1]$, $\lim_{L \rightarrow
\infty} \tilde{P}_f(B,L) = 0$ for all $B \in (0,\tilde{q}_c)$,
$\lim_{L \rightarrow \infty} \tilde{P}_m(B,L;p_d) = 0$ for all $B
\in (0,\tilde{q}_c-\tilde{q}_d)$, and $\lim_{L \rightarrow \infty} \tilde P_m(B,L;p_d) = 1$ for all $B
\in (\tilde{q}_c-\tilde{q}_d,\tilde{q}_c)$.
\end{lemma}

\begin{IEEEproof}
We use the same approach as in the proof of Lemma 1. When every node
transmits with probability $p_c$, the idle slot ratio converges almost
surely to $\tilde{q}_c$ as $L$ goes to infinity, which implies that the false
punishment probability goes to zero for all $B > 0$. When
there is exactly one node transmitting with probability $p_d$, the
idle slot ratio converges almost surely to $\tilde q_d$ as $L$ goes to infinity. Hence, if
$\tilde q_d < \tilde q_c - B$ (resp. $\tilde q_d > \tilde q_c - B$), the miss detection probability goes to
zero (resp. one).
\end{IEEEproof}

Lemma 2 gives a sufficient condition on the idle slot ratio test to apply Proposition 5.

\begin{proposition}
Suppose that $B \in (0,\tilde{q}_c-\tilde{q}_d)$. For any
$\delta > 0$, there exist $L$ and $M$ such that
$\tilde{\sigma}^r(B,L,M)$ is DP against $\tilde{\sigma}^d$ and $\delta$-PO.
\end{proposition}

\begin{IEEEproof}
The proposition follows from Lemma 2 and Proposition 5.
\end{IEEEproof}
Proposition 7 states that for given $p_d \in (p_c,1]$, we can always construct
a protocol based on the idle slot ratio test that is DP against $\tilde{\sigma}^d$ and achieves an
arbitrarily small efficiency loss by choosing $B$ such that $0 < B
< \tilde{q}_c-\tilde{q}_d = (p_d-p_c)(1-p_c)^{N-1}$. As in the case of the ACK ratio test,
we have a wider range of $B$ that renders deviation-proofness as $p_d$ is larger.


We have considered deviation strategies that prescribe a constant transmission
probability in a review phase.
We now consider the case where a deviating node can use any strategy in $\Sigma_i$,
which includes strategies that adjust transmission probabilities depending on
the signals obtained in the current review phase.
The following theorem shows that we can construct a protocol based on the idle slot
ratio test that is deviation-proof and near-optimal.

\begin{theorem}
For any $\epsilon > 0$ and $\delta > 0$, there exist $B$, $L$, and $M$
such that $\tilde{\sigma}^r(B,L,M)$ is $\epsilon$-NE and $\delta$-PO.
\end{theorem}

\begin{IEEEproof}
The proof is relegated to Appendix B.
\end{IEEEproof}
The interpretation of $\epsilon$ and $\delta$ as performance requirements
as well as the trade-off between performance and implementation cost,
as discussed following Theorem 2, is still valid in the case of public signals.

{\it Remark: Protocols with sliding windows.}
Suppose that more than $L(\tilde{q}_c-B)$ idle slots have occurred
before the end of a review phase.
Then, a deviating node, knowing that a punishment will not occur
regardless of the outcome in the remaining slots of the review
phase, can increase its transmission probability for
the remainder of the review phase to obtain a payoff gain. We can
make a protocol based on a review strategy robust to such a
manipulation by having sliding windows for review phases.
In a review strategy with sliding windows, a review phase begins in each slot unless there is
a new or ongoing punishment. Once the idle slot ratio test based on
the recent $L$ signals fails, a review stops and punishment
occurs for $M$ slots. Once a punishment phase ends, a review phase
begins in each slot until another punishment occurs.
A detailed analysis of this protocol is left for future research.

\subsection{Numerical Results}

We provide numerical results to demonstrate the findings on DP protocols
with public signals. Again, we consider a network with $N=5$ and $p_c=1/N=0.2$ while
varying $p_d$ and the protocol parameters.

\begin{figure*}
\centering
\vspace{-0.1cm}
\subfloat[]{\includegraphics[width=9cm, height=5cm]{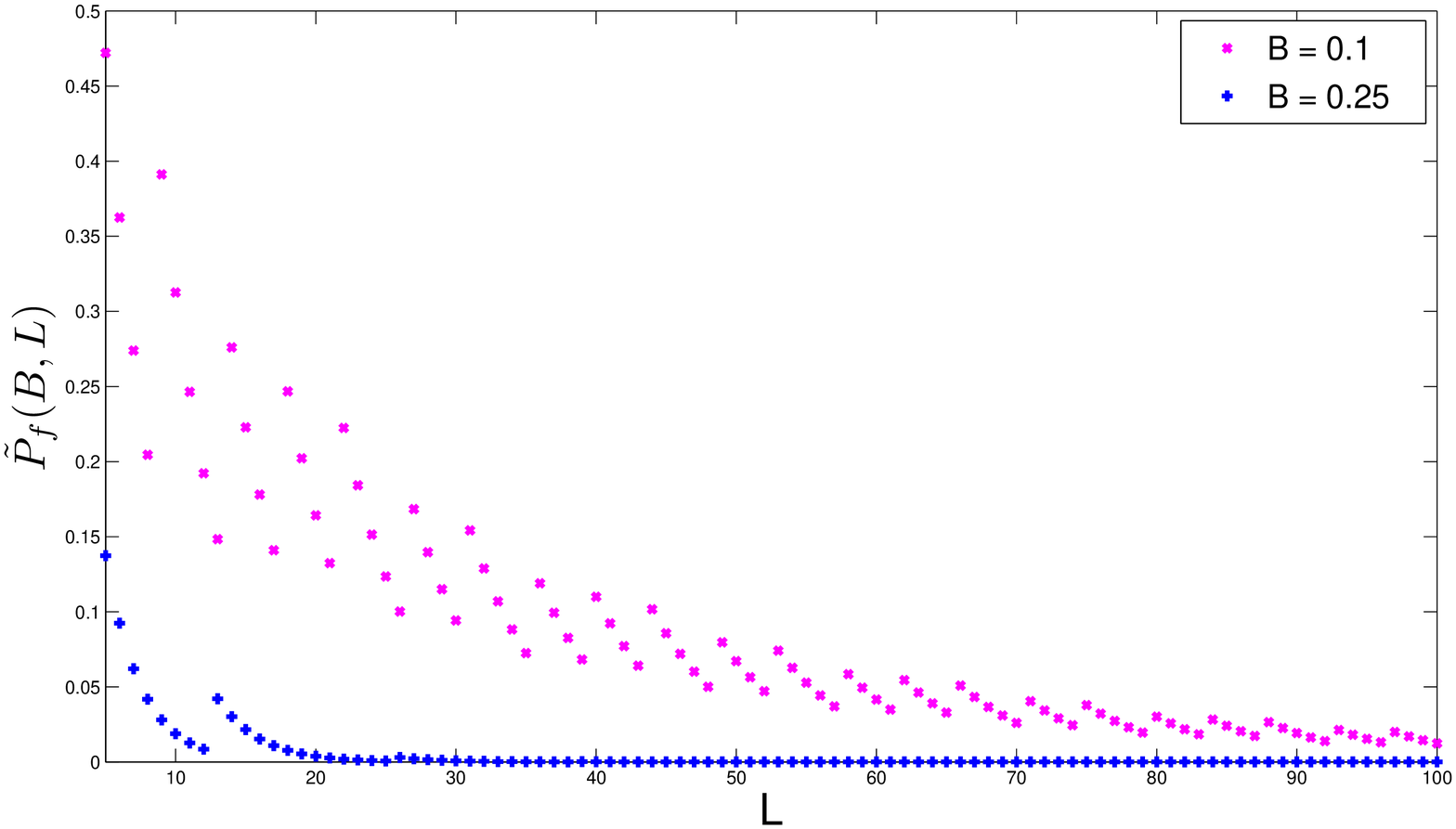}}
\subfloat[]{\includegraphics[width=9cm, height=5cm]{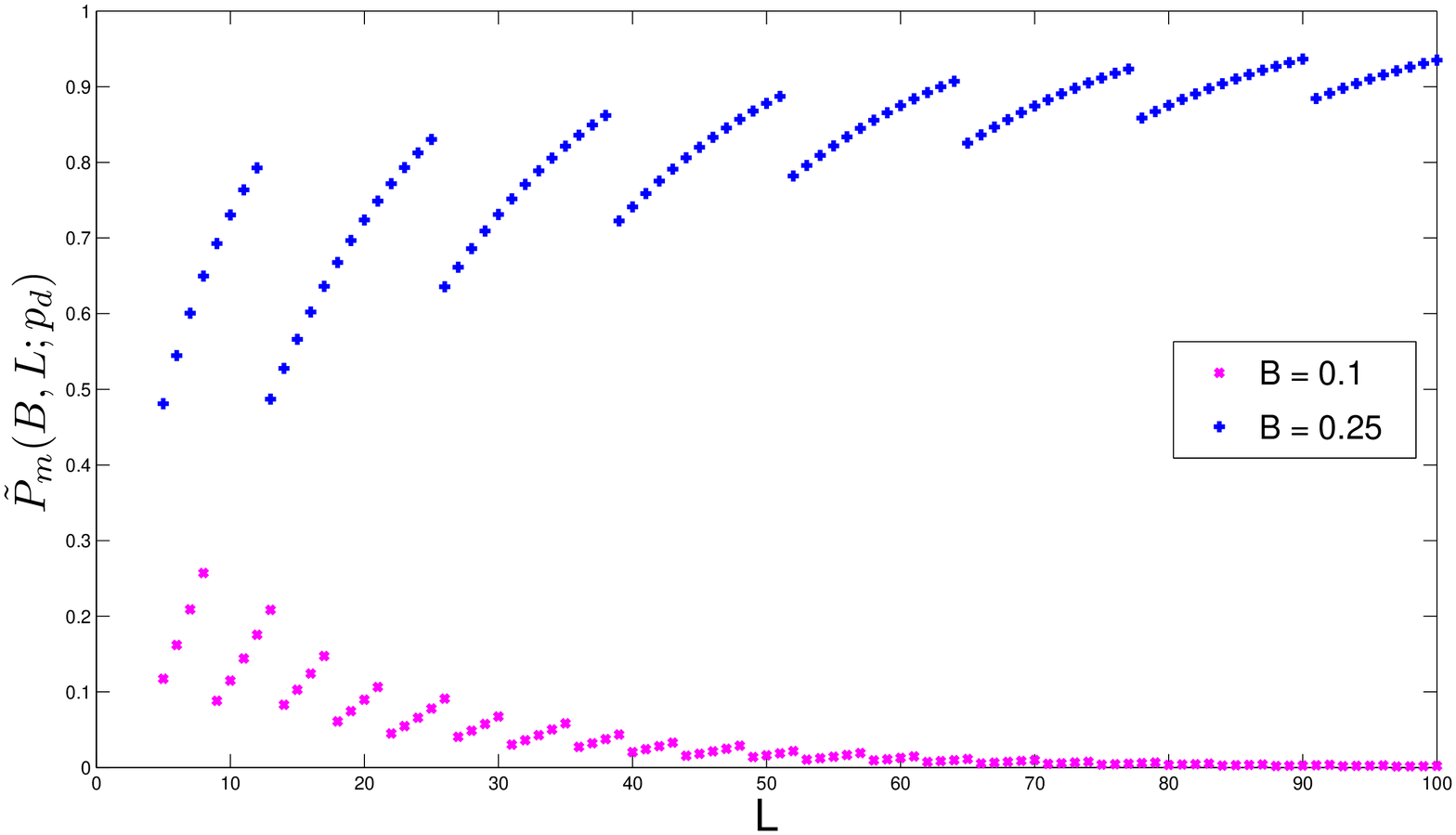}}
\caption{$\tilde P_f(B,L)$ and $\tilde P_m(B,L;p_d)$ versus the length of a review phase $L$ when $p_d = 0.7$.}
\label{Fig6}
\end{figure*}

Fig.~\ref{Fig6} plots $\tilde P_f$ and $\tilde P_m$ against the length of a review phase $L$ for $B = 0.1$ and $0.25$ when $p_d = 0.7$. As in the case of private signals, $\tilde P_f$ tends to decrease with $L$ and approaches zero for large $L$. Also, $\tilde P_f$ is smaller for a larger margin of error $B$. Note that the upper threshold for $B$ to yield $\lim_{L \rightarrow \infty} \tilde{P}_m(B,L;p_d) = 0$ in Lemma 2 is $\tilde q_c - \tilde q_d = 0.2048$. We can see that when $B$ is larger than this threshold, $\tilde P_m$ tends to increases with $L$ and approaches 1 for large $L$. On the contrary, when $B$ is smaller than the threshold, $\tilde P_m$ approaches zero for large $L$, making the test asymptotically perfect.

\begin{figure*}
\centering
\subfloat[]{\includegraphics[width=9cm, height=5cm]{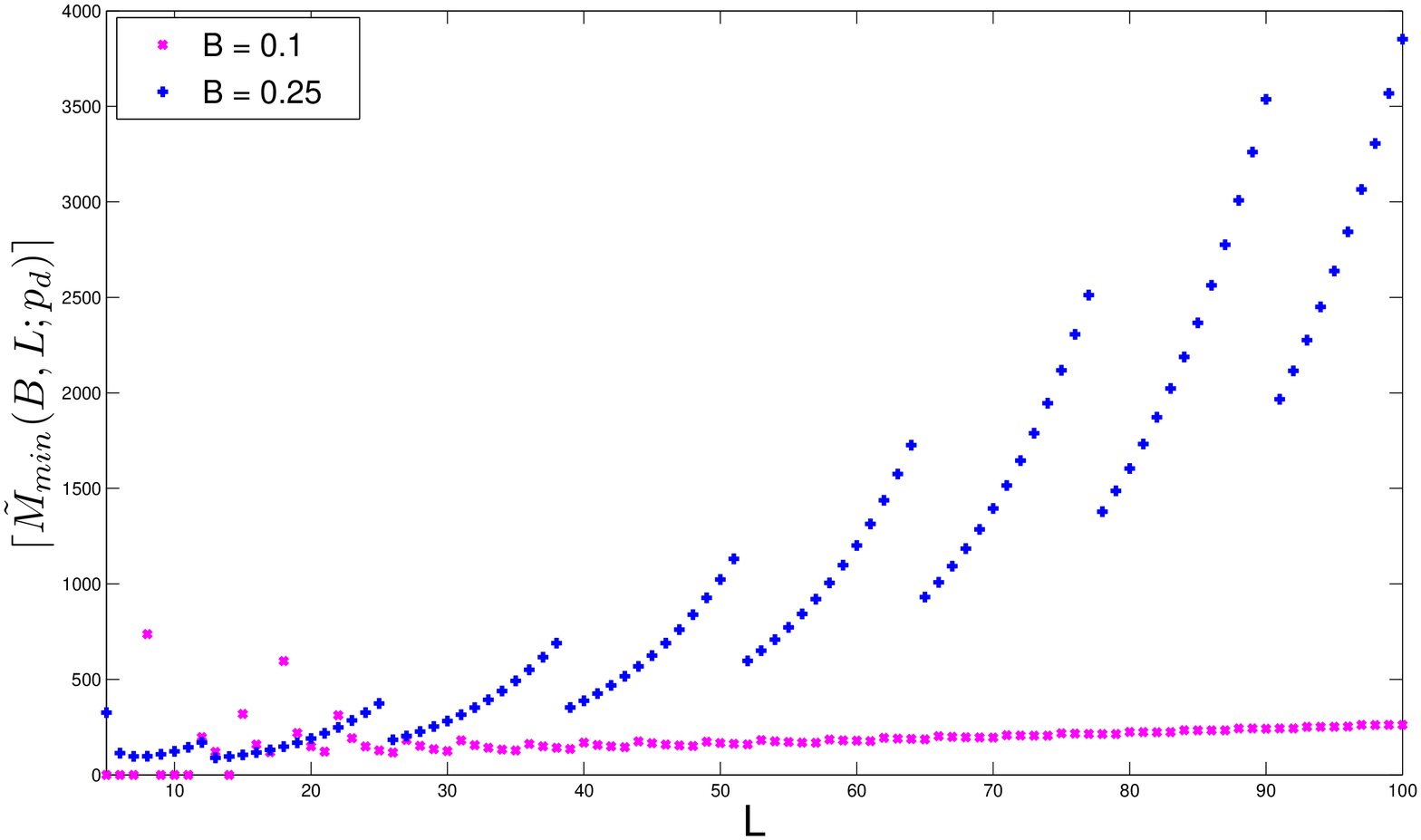}}
\subfloat[]{\includegraphics[width=9cm, height=5cm]{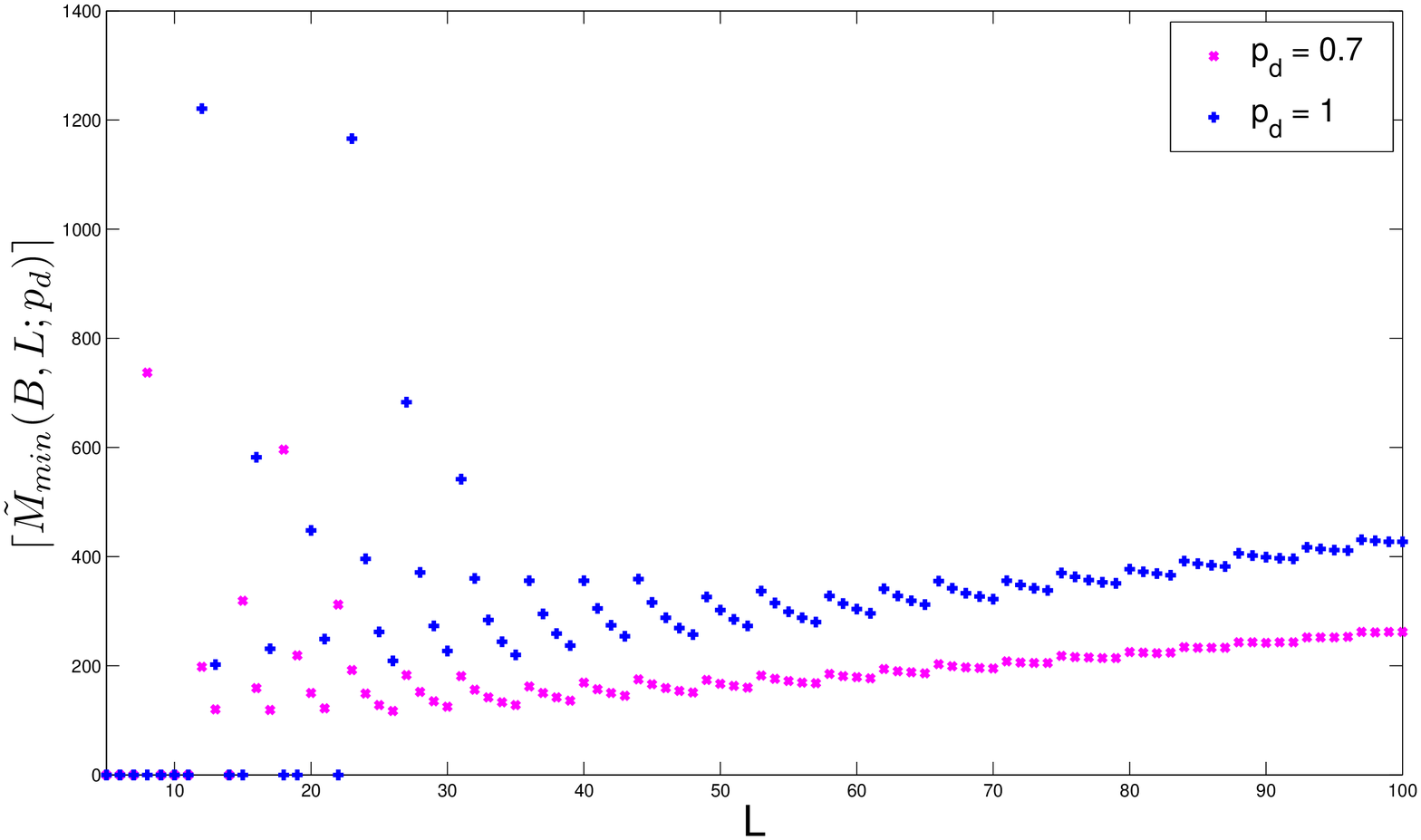}}
\caption{The minimum length of a punishment phase $\lceil
\tilde{M}_{\rm min}(B,L;p_d) \rceil$ versus the length of a review phase $L$: (a) $p_d = 0.7$, and (b) $B = 0.1$.}
\label{Fig7}
\vspace{-0.35cm}
\end{figure*}

\begin{figure*}
\centering
\subfloat[]{\includegraphics[width=9cm, height=5cm]{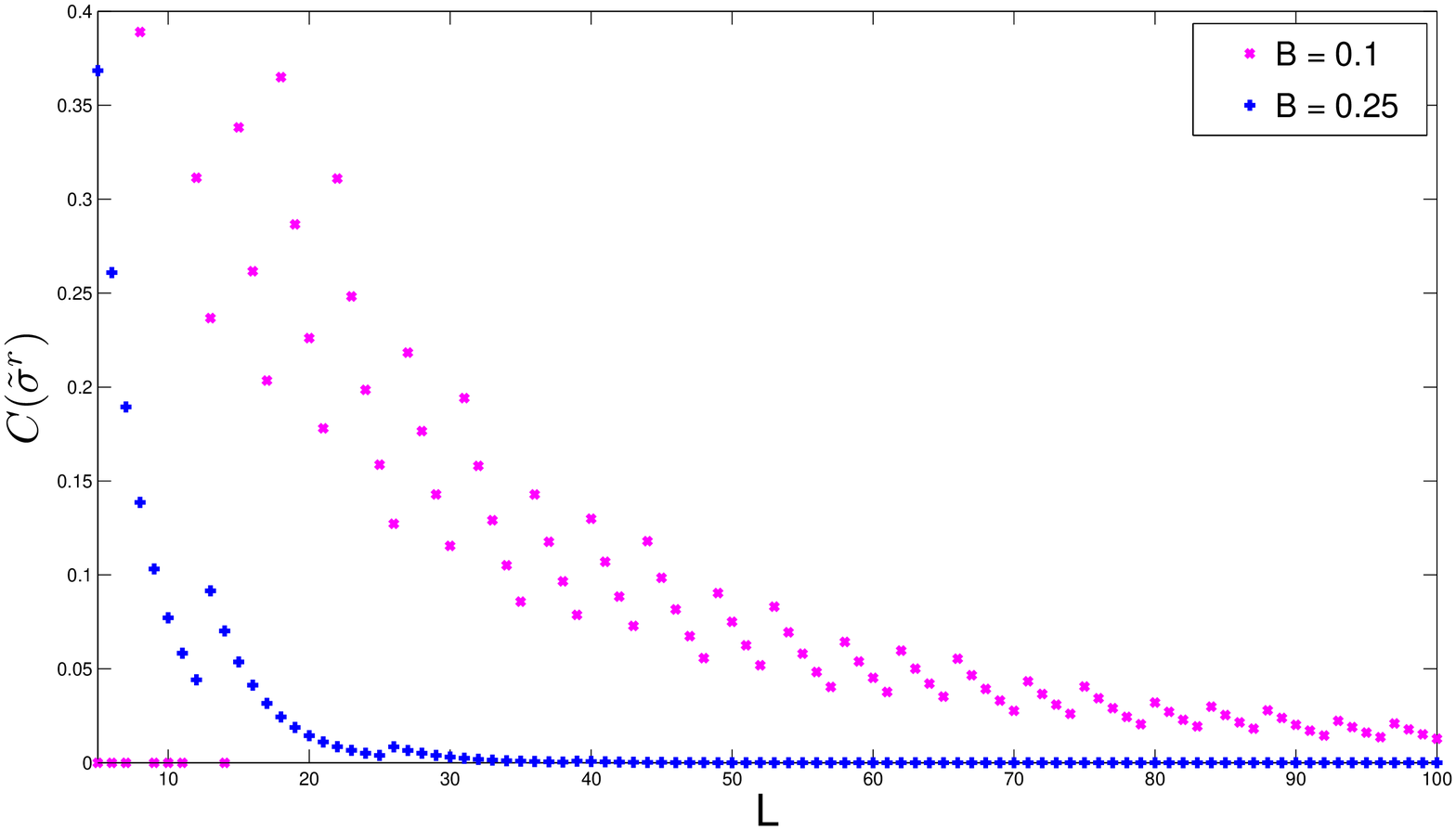}}
\subfloat[]{\includegraphics[width=9cm, height=5cm]{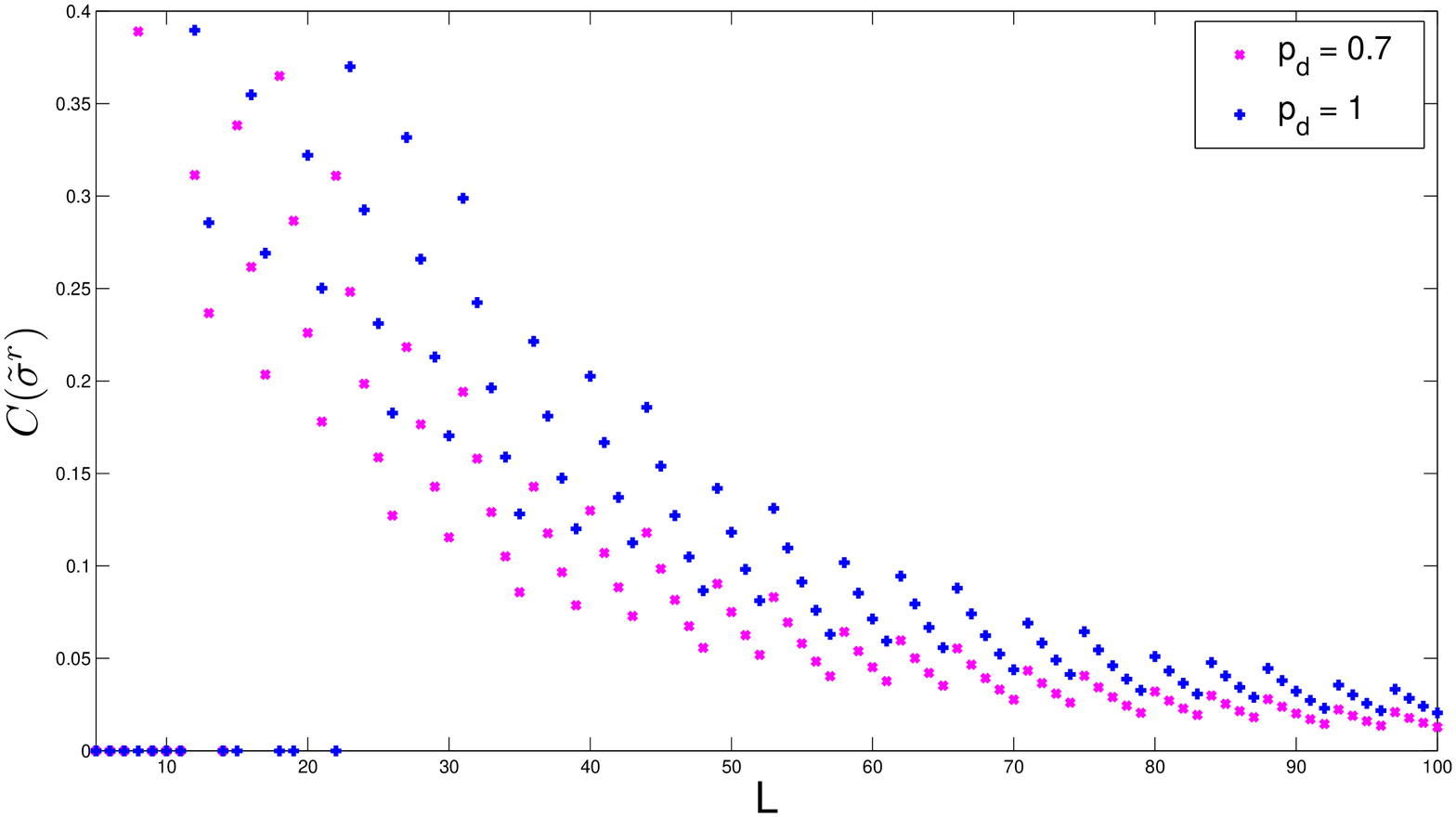}}
\caption{Efficiency loss $C(\tilde \sigma^r)$ versus the length of a review phase $L$: (a) $p_d = 0.7$, and (b) $B = 0.1$.}
\label{Fig8}
\vspace{-0.35cm}
\end{figure*}

Fig.~\ref{Fig7} plots the minimum length of a reciprocation phase $\lceil \tilde M_{\rm min}(B,L;p_d)\rceil$ to have a DP protocol as a function of the length of a review phase $L$. We can see that for fixed $p_d = 0.7$, a longer reciprocation phase is needed for larger $B$, except when $L$ is small, and that DP protocols cannot be constructed with some small values of $L$ when $B=0.1$ (displayed as $\lceil \tilde M_{\rm min}(B,L;p_d)\rceil = 0$). Also, when $B = 0.1$, a longer reciprocation phase is needed for $p_d = 1$ than for $p_d = 0.7$. The efficiency loss of DP protocols with the minimum length of a reciprocation phase is shown in Fig.~\ref{Fig8}. We can see that larger $B$ results in smaller efficiency loss, because $\tilde P_f$ is smaller for larger $B$ as shown in Proposition 6. Also, efficiency loss approaches zero as $L$ becomes large, which is consistent with Proposition 7.

%

\section{Extension to a CSMA/CA Network with Selfish Nodes}

In this section, we discuss how the proposed protocols based on a review strategy can be 
modified for a CSMA/CA network. 
In \cite{Cagalj}, the authors consider a CSMA/CA network in which a selfish node uses
a fixed contention window size.
They show a discrepancy between NE and Pareto optimum. The contention window size of each node
at the unique PO outcome is denoted by $W^*$, which results in a transmission probability 
$p_c = 2/(W^*+1)$. The optimal payoff $u^{\rm PO}$, i.e., the throughput at Pareto optimum, 
can be computed using Eq. (1) of \cite{Cagalj}, based on the model of \cite{Bianchi}.

A review strategy for a CSMA/CA network can be described as follows, assuming 
private signals (i.e., sensing information is private). At the beginning, nodes
are synchronized to start a review phase. In a review phase, which lasts for $L$ time period,
each node sets its window size at $W^*$. After a review phase, each node computes its
actual throughput, denoted by $\tau_i$, and compares it with $u^{\rm PO}$, the expected throughput
when no node has deviated from $W^*$. A deviating node chooses its window size $W^d$ smaller than $W^*$ in order to 
increase its transmission probability from $p_c$ and thus to obtain a higher throughput.
Since a deviation decreases the throughput of the well-behaved nodes, we can design a test
such that the test performed by node $i$ is passed if and only if $\tau_i > u^{\rm PO} - B$ for some constant $B
\in (0, u^{\rm PO})$. If the test of node $i$ is passed, node $i$ moves to a cooperation phase
during which it continues to set its window size at $W^*$. Otherwise, it moves to a
punishment phase during which it sets its window size at the minimum value 1. A reciprocation
phase lasts for $M$ time period, and a new review phase begins after a reciprocation phase.

As in a slotted Aloha network, $\tau_i$ converges almost surely to $u^{\rm PO}$
as $L$ goes to infinity, and thus the proposed test can be made asymptotically
perfect by choosing an appropriate value of $B$. Hence, when window sizes take discrete 
values, we can construct a protocol that is DP against any constant deviation strategy 
and achieves a small efficiency loss, following a similar approach to Theorem 2.
We omit the details due to lack of space.

\section{Conclusion}

It is well-known that the decentralized operation of multiple access communication systems with selfish nodes
often results in an inefficient use of a shared medium.
To overcome this problem, we have proposed new classes of slotted MAC protocols that are robust to
selfish manipulation while achieving near-optimality.
The proposed protocols are based on the idea of a review strategy in the theory of repeated
games. With the proposed protocols, nodes perform a statistical test to determine whether a deviation
has occurred and trigger a punishment when they conclude so. We have provided conditions under which
we can design deviation-proof protocols with a small efficiency loss and illustrated
the results with particular statistical tests.  Our framework and design methodology are not limited to
multiple access communications. They can be applied to other networking and communication
scenarios in which agents obtain imperfect signals about the decisions of other agents
and a deviation influences the distribution of signals.

\appendices

\section{Proof of Theorem 2}
Choose arbitrary $\epsilon > 0$ and $\delta > 0$. Define $p_{\epsilon} \triangleq p_c + \epsilon/(1-p_c)^{N-1}$. Note that $p_{\epsilon}$ is
the minimum deviation probability with which a deviating node gains at least $\epsilon$ in a slot
when other nodes transmit with probability $p_c$. Choose $B \in (0, \epsilon/(N-1))$. Note that $q_c - q_d \geq \epsilon/(N-1)$
for all $p_d \in [p_{\epsilon},1]$.
Define
\begin{align*}
\hat{g}(B,L) \triangleq& \ \left(1-P_{f}(B,L)\right)^{\frac{N-1}{N}} - (1-p_c) \left(1-P_f(B,L)\right) \\
& - P_m(B,L;p_{\epsilon}). 
\end{align*}
Since $P_m(B,L;p_d)$ is non-increasing in $p_d$, we have $g(B,L;p_d) \geq \hat{g}(B,L)$
for all $p_d \in [p_{\epsilon},1]$, where $g(B,L;p_d)$ is defined in \eqref{pun3}. Also, by Lemma 1, we have $\lim_{L \rightarrow
\infty} P_f(B,L) = 0$ and $\lim_{L \rightarrow \infty} P_m(B,L;p_{\epsilon}) = 0$.
Therefore, $\lim_{L \rightarrow \infty} \hat{g}(B,L) = p_c$, and thus
there exists $L_1$ such that $g(B,L;p_d) > 0$ for all $p_d \in [p_{\epsilon},1]$,
for all $L \geq L_1$.
Define $\hat{M}(L) \triangleq \lceil (1-p_c)L/\hat{g}(B,L) \rceil$. Since $\hat{M}(L) \geq
M_{\rm min}(B,L;p_d)$ for all $p_d \in [p_{\epsilon},1]$, protocol $\sigma^r(B,L,\hat{M}(L))$ is DP
against all constant strategies using $p_d \in [p_{\epsilon},1]$, for all $L \geq L_1$.

Since $C(\sigma^r)$ is non-decreasing in $M$, we have
\begin{align*}
0 \leq C(\sigma^r(B,L,\hat{M}(L))) \leq \frac{N \left((1-p_c)L/\hat{g}(B,L) + 1\right)}{L+\left((1-p_c)L/\hat{g}(B,L) + 1\right)} \\
\times (1-p_c)^{N-1} \left[ p_c P_{f} - (1-P_f)^{\frac{N-1}{N}} + (1 -P_f) \right].
\end{align*}
Therefore, $\lim_{L \rightarrow \infty} \mathcal{P}_s(\sigma^r) = 0$, and
there exists $L_2$ such that $C(\sigma^r) < \delta$
for all $L \geq L_2$. Choose $L \geq \max \{L_1, L_2\}$ and
$M = \hat{M}(L)$. Then $\sigma^r(B,L,M)$ is DP
against all constant strategies using $p_d \in [p_{\epsilon},1]$ and satisfies $C(\sigma^r) < \delta$.
Finally, note that the payoff gain from deviating to
a constant strategy using $p_d \in [0,p_{\epsilon})$ is bounded above by $\epsilon$.
Hence, $\sigma^r(B,L,M)$ is robust $\epsilon$-DP and $\delta$-PO. This completes the proof.

\section{Proof of Theorem 4}
Consider the problem of a deviating node maximizing its payoff given that all the other
nodes use a review strategy $\tilde{\sigma}^r(B,L,M)$, i.e., $\max_{\sigma \in \Sigma_i}
U(\sigma; \tilde{\sigma}^r)$. We can define a state space with total $L(L+1)/2 + M$
states, where a state is a pair consisting of the slot position and the number of idle slots since the beginning
of the current review phase in the case of a review phase while it is the slot position in the case of a punishment phase.
By the principle of dynamic programming, we can obtain a stationary optimal strategy,
denoted by $\sigma^*$. Let $p_t$ be the expected value of the transmission probability of a node using $\sigma^*$
in slot $t$ of a review phase (conditional on null history) when other nodes follow $\tilde{\sigma}^r$. Let $I_t = \chi\{z^t = 0\}$.
Since $E[I_t] = (1-p_t)(1-p_c)^{N-1}$, we have
\beqa
U(\sigma^*; \tilde{\sigma}^r) \!\!\!&=&\!\!\! \frac{(1-p_c)^{N-1} \sum_{t=\tau+1}^{\tau+L} p_t}{L+ P^*_f M} \nonumber \\
\!\!\!&=&\!\!\!  \frac{L(1-p_c)^{N-1} - E \left[ \sum_{t=\tau+1}^{\tau+L} I_t\right]}{L+P^*_f M}, \label{eq:usig}
\eeqa
where $\tau + 1$ is the first slot of a review phase and $P^*_f$
is the punishment probability when the deviating node uses $\sigma^*$, i.e., $P^*_f =  \Pr\left\{\sum_{t=\tau+1}^{\tau+L} I_t \leq L(\tilde{q}_c - B) \right\}$.
Since $\sum_{t=\tau+1}^{\tau+L} I_t \geq 0$, using Markov's inequality, we have
\begin{align} \label{eq:eit}
E \left[\sum_{t=\tau+1}^{\tau+L} I_t \right] \geq (1-P_f^*)L(\tilde{q}_c - B).
\end{align}
Combining \eqref{eq:usig} and \eqref{eq:eit}, we obtain
\begin{align} \label{eq:usig2}
U(\sigma; \tilde{\sigma}^r) \leq
\frac{L q_c + P^*_f (1-p_c)^{N} L + (1-P^*_f) BL}{L+P^*_f M}
\end{align}
for all $\sigma \in \Sigma_i$.

Choose arbitrary $\epsilon > 0$ and $\delta >0$.
Following \cite{RoyRadner}, we relate the choice of $M$ and $B$ to $L$ as follows:
\begin{align} 
&B = \beta L^{\rho-1}, \quad \beta > 0, \quad \frac{1}{2} < \rho < 1  \nonumber\\
&M = \mu L, \quad \mu > 0  \nonumber
\end{align}
Fix $\beta$, $\rho$, and $\mu$ such that
$\beta > 0$, $1/2 < \rho < 1$, and $\mu > N-1$.
By Chebychev's inequality,
\begin{align} \label{false}
\tilde{P}_f(B,L) \leq \frac{\tilde{q}_c (1-\tilde{q}_c)}{B^2 L} =
\frac{\tilde{q}_c (1-\tilde{q}_c)}{\beta^2 L^{2\rho-1}}.
\end{align}
Also, note that
\begin{align} \label{newPoS}
C(\tilde{\sigma}^{r}) = \frac{N \tilde{P}_f M q_c}{L+\tilde{P}_f M}
= \frac{N \tilde{P}_f \mu q_c}{1+\tilde{P}_f \mu}.
\end{align}
Since $\tilde{P}_f(B,L)$ in (\ref{false}) converges to zero as $L$ goes to infinity,
we can achieve an arbitrarily small efficiency loss in (\ref{newPoS}) by choosing sufficiently large $L$.
In other words, for any $\delta > 0$, there exists $L'_{\delta}$ such that
$C(\tilde{\sigma}^{r}) < \delta$ for all $L \geq L'_{\delta}$.
With $\mu > N-1$, the upper bound on the deviation payoff in \eqref{eq:usig2}
\begin{align} 
\frac{q_c + P^*_f (1-p_c)^{N} + (1-P^*_f) \beta L^{\rho-1}}{1+ P^*_f \mu} \nonumber
\end{align}
is decreasing in $P^*_f$. Thus, the deviation payoff is bounded above by $q_c+\beta L^{\rho-1}$.

Choose $L$ such that
\begin{align}
L \geq \max \left\{ L'_{\delta}, L'_{N\epsilon/2}, \Bigl(\frac{2\beta}{\epsilon}\Bigr)^{\frac{1}{1-\rho}} \right\}.  \nonumber
\end{align}
Since $L \geq L'_{N\epsilon/2}$, we have
\begin{align} \label{eq:ulower}
q_c - \frac{\epsilon}{2} <U(\tilde{\sigma}^r; \tilde{\sigma}^r) \le U(\sigma^*; \tilde{\sigma}^r).
\end{align}
Since $L \geq \left(2\beta/\epsilon \right)^{1/(1-\rho)}$, we have
\begin{align} \label{eq:upper}
U(\sigma^*; \tilde{\sigma}^r) \le q_c+\beta L^{\rho-1} \le q_c + \epsilon/2.
\end{align}
Then, by \eqref{eq:ulower} and \eqref{eq:upper}, we obtain an upper bound on the deviation gain as
\begin{align}
U(\sigma^*; \tilde{\sigma}^r) - U(\tilde{\sigma}^r; \tilde{\sigma}^r) \leq
\epsilon, \nonumber
\end{align}
which proves that $\tilde{\sigma}^r(B,L,M)$ is an $\epsilon$-NE.
Lastly, since $L \geq L'_{\delta}$, we have $C(\tilde{\sigma}^{r}) < \delta$, and thus $\tilde{\sigma}^r(B,L,M)$ is $\delta$-PO.

\end{document}